\documentclass[aps,prx,showpacs,floatfix,twocolumn,superscriptaddress,nofootinbib,longbibliography]{revtex4-2}
\usepackage{graphicx}
\usepackage{bm,amsmath}
\usepackage{dcolumn}
\usepackage{amsfonts,amssymb}
\usepackage[version=4]{mhchem}
\usepackage{diagbox}
\usepackage{hyperref}
\usepackage{subfigure}
\usepackage[mathlines]{lineno}
\begin{document}
\title{Quantum phase transition  and composite  excitations of antiferromagnetic spin trimer chains in a magnetic field }

\author{Jun-Qing Cheng}

\affiliation{State Key Laboratory of Optoelectronic Materials and Technologies, Center for Neutron Science and Technology, Guangdong Provincial Key
Laboratory of Magnetoelectric Physics and Devices, School of Physics, Sun Yat-Sen University, Guangzhou 510275, China}
\affiliation{School of Physical Sciences, Great Bay University, 523000, Dongguan, China}
\affiliation{Great Bay Institute for Advanced Study, Dongguan 523000, China}

\author{Zhi-Yao Ning}

\affiliation{State Key Laboratory of Optoelectronic Materials and Technologies, Center for Neutron Science and Technology, Guangdong Provincial Key
Laboratory of Magnetoelectric Physics and Devices, School of Physics, Sun Yat-Sen University, Guangzhou 510275, China}

\author{Han-Qing Wu}
\email{wuhanq3@mail.sysu.edu.cn}
\affiliation{State Key Laboratory of Optoelectronic Materials and Technologies, Center for Neutron Science and Technology, Guangdong Provincial Key
Laboratory of Magnetoelectric Physics and Devices, School of Physics, Sun Yat-Sen University, Guangzhou 510275, China}

\author{Dao-Xin Yao}
\email{yaodaox@mail.sysu.edu.cn}
\affiliation{State Key Laboratory of Optoelectronic Materials and Technologies, Center for Neutron Science and Technology, Guangdong Provincial Key
Laboratory of Magnetoelectric Physics and Devices, School of Physics, Sun Yat-Sen University, Guangzhou 510275, China}

\date{\today}

\begin{abstract}
	Motivated by recent advancements in theoretical and experimental studies of the high-energy excitations on an antiferromagnetic trimer chain, we numerically investigate the quantum phase transition and composite  dynamics in this system by applying  a magnetic field.  The numerical methods we used include the exact diagonalization, density matrix renormalization group, time-dependent variational principle, and  cluster perturbation theory. From calculating the entanglement entropy, we have revealed   the phase diagram which includes the XY-I, $1/3$ magnetization plateau, XY-II, and ferromagnetic phases.  Both the critical  XY-I and XY-II phases are  characterized by the conformal field theory with a central charge $c \simeq 1$. By analyzing the dynamic spin structure factor, we elucidate the distinct features of spin dynamics across different phases. In the regime with weak intertrimer interaction, we identify the intermediate-energy and high-energy modes in the  XY-I and $1/3$ magnetization plateau phases as internal trimer excitations,  corresponding to the propagating of doublons and quartons, respectively. Notably, applying a magnetic field splits the high-energy spectrum into two branches, labeled as the upper quarton and lower quarton. Furthermore, we  explore the spin dynamics of a frustrated trimerized  model  closely related to the quantum magnet \ce{Na_2Cu_3Ge_4O_12}. In the end, we extend our  discuss  on the possibility of the quarton Bose-Einstein condensation in the trimer systems.  Our results are expected to be further verified through the  inelastic neutron scattering and resonant inelastic X-ray scattering,  and also provide valuable insights for  exploring  high-energy exotic excitations.
\end{abstract}

\maketitle

\section{\label{sec:level1} Introduction}

Understanding  the  profound physical nature of the strongly correlated many-body systems is a  challenging and fascinating task in modern condensed-matter physics. Among the various physical properties, magnetic excitation plays a crucial role in  understanding the magnetic structures of quantum materials and can be studied both theoretically and experimentally \cite{Mikeska2004,karbach97,tennant93,lake13,PhysRevLett.104.237207,PhysRevLett.106.157205,RIXS2018NC,PhysRevLett.102.037203, wang2018experimental,PRL2019ZheWang,cheng2022,bera2022emergent,do2023understanding,han2023weak,PhysRevLett.105.247001,Zhou2013,Ishii2014,Song2021,SAC2,dalla2015fractional,PhysRevB.108.224418,chang2023magnon}. 
In particular,  the strong quantum fluctuations in low-dimensional systems give rise to  a variety of exotic ground states and excitations, such as the Luttinger liquid and spinon excitation, which   have attracted significant interest \cite{Mikeska2004,karbach97,tennant93,lake13,PhysRevLett.104.237207,PhysRevLett.106.157205,RIXS2018NC,PhysRevLett.102.037203, wang2018experimental,PRL2019ZheWang,cheng2022,bera2022emergent}.  Quasi one-dimensional (1D) magnetic materials can be effectively described by the Heisenberg spin chain, and its various extensions have been extensively investigated.
 For example, 
 the gapless two-spinon continuum \cite{karbach97}  has been observed through inelastic neutron scattering in quasi 1D material \ce{KCuF_3} \cite{tennant93,lake13} and   the
frustrated ferromagnetic spin-$1/2$ chain compound \ce{LiCuVO_4} \cite{PhysRevLett.104.237207}. Multi-spin excitations can be detected using the 
resonant inelastic X-ray scattering (RIXS) technique  in the  material \ce{Sr_2CuO_3} \cite{PhysRevLett.106.157205,RIXS2018NC}. Furthermore, the high-energy string excitations have  been proposed as the dominant excitations in the isotropic Heisenberg antiferromagnet based on the Bethe ansatz \cite{PhysRevLett.102.037203}, and have recently  been observed in an antiferromagnetic Heisenberg–Ising chain compounds \ce{SrCo_2V_2O_8} and \ce{BaCo_2V_2O_8} under strong longitudinal magnetic fields using the high-resolution terahertz spectroscopy  \cite{wang2018experimental,PRL2019ZheWang}.

Besides the uniform spin chains, 
quantum materials often exhibit structures that consist of more than one spin per unit cell, resulting in more rich magnetic properties. Among that,  ladder systems are well-studied examples of quasi-1D systems with more spins in a unit cell, where the gapless or gapped  excitation spectrum  depends on whether the  rungs contain an odd or even number of $S=1/2$ spins, respectively \cite{dagotto96}. This behavior is analogous
to the Haldane's conjecture regarding spin chains with half-odd-integer or integer spins \cite{haldane1983}. Among the experimental realizations, the two-leg ladder
compound \ce{(C_7H_{10}N)_2CuBr_4} is noteworthy due to the excellent agreement between its inelastic neutron scattering spectrum and the dynamic spin structure factor derived from the model calculations \cite{schmidiger13}. Besides the ladder spin system, spatial inhomogeneous is another way to enlarge the unit cell with more spins. For instance, the model featuring two-spin unit cells with alternating couplings $J_1 - J_2$, called alternating or dimerized Heisenberg chain,
has been well studied. A gap appears in the spectrum if $J_1 \not= J_2$, as this modulation causes the spinons of the uniform Heisenberg antiferromagnetic chain ($J_1=J_2$) to be confined
into triplons that can be considered as weakly bound of spinons  when $J_2 \approx J_1$ \cite{doretto09}. Extended to trimerized system with
 three-spin unit cells and repeated couplings $J_1 - J_1 - J_2$, the so-called trimerized Heisenberg chain is less studied and would exhibit very different magnetic excitations due to odd number of spins in a unit cell \cite{cheng2022}. Moreover, this trimerized structure has been observed in real materials like \ce{A_3Cu_3(PO_4)_4}
(\ce{A=Ca, Sr, Pb}) \cite{PhysRevB.71.144411,drillon19931d,belik2005long,PhysRevB.76.014409,PhysRevB.102.035137,PhysRevB.105.134423}, \ce{(C_5H_11NO_2)_2.3CuCl_2.2H_2O}
\cite{hasegawa2012magnetic}, and \ce{Ba_4Ir_3O_10} \cite{cao2020quantum,PhysRevLett.129.207201}.   In the  iridate \ce{Ba_4Ir_3O_10}, where three-spin unit cells form  layered trimers,  fractional spinon excitation has  have been observed in the RIXS experiments \cite{cao2020quantum,PhysRevLett.129.207201}.

In our previous work \cite{cheng2022}, we have investigated the spin dynamics of  trimer chain characterized by repeated couplings
$J_1 - J_1 - J_2$, whereas the intratrimer $J_1$ is larger than intertrimer $J_2$. Our findings show  that the low-energy excitation corresponds to the two-spinon continuum can be well described by the uniform Heisenberg model of effective trimer block spins \cite{cheng2022}. 
Most interestingly, some new composite excitations of the
novel  quasi-particles, known as doublons and quartons, have been predicted in  the intermediate-energy and high-energy spectra, respectively, and have been subsequently  confirmed in the 
inelastic neutron scattering  measurements on \ce{Na_2Cu_3Ge_4O_12} \cite{bera2022emergent}. As $J_2/J_1 \rightarrow 1$, the doublons and quartons lose their identities and fractionalize into the conventional two-spinon continuum. We have also extended the doublon and quarton excitations to 2D trimer systems \cite{chang2023magnon}. Even though we have a clear understanding these exotic excitations, how to control these excitations using magnetic field is still an interesting topic.

In this paper, we want to study the effect of magnetic field on the doublons and quartons in the trimer chain illustrated in Fig.~\ref{diagram}(a). To achieve this, we employ various techniques, including  exact diagonalization (ED), density matrix renormalization group (DMRG) \cite{PhysRevLett.69.2863,PhysRevLett.93.076401,SCHOLLWOCK201196}, time-dependent variational principle (TDVP) \cite{PhysRevLett.107.070601, PhysRevB.94.165116},  and cluster perturbation theory (CPT) \cite{PhysRevB.48.418,PhysRevLett.84.522,RevModPhys.77.1027,PhysRevB.98.134410} to investigate the excitation spectra of trimer chain under the  magnetic field.  By mapping the entanglement entropy onto the parameter space,  we identify the  XY-I, the $1/3$ magnetization plateau, the XY-II and the ferromagnetic phases. In the gapless XY-I and XY-II phases, both  central charges $c \simeq 1$ indicate that these two  phases   are well described by the  conformal field theory.  More importantly,  we  investigate the intermediate-energy and high-energy excitations for small $g$  in the XY-I and $1/3$ magnetization plateau  phases. Our analysis demonstrates that  the intermediate-energy and high-energy modes are primarily governed by the internal trimer excitations,  referred to as the doublons and quartons, respectively. Furthermore, these features of the excitation spectra can also be observed in the spin chain with a trimer structure that is closely associated with the quantum magnet \ce{Na_2Cu_3Ge_4O_12} \cite{bera2022emergent}. The magnetic field drives the lower quarton toward  zero energy,  suggesting the potential for observing  the  magnetic-field-induced  quarton 
Bose-Einstein condensation (BEC) in the quantum magnet \ce{Na_2Cu_3Ge_4O_12}.
Our results may facilitate further exploration of the high-energy spin excitation mechanisms in other systems containing clusters with odd spins.


\section{Results}

\subsection{Model}

The Hamiltonian of the spin-$1/2$ antiferromagnetic trimer chain subjected to  a longitudinal magnetic field  reads
\begin{eqnarray}
	\mathcal{H}&=&\sum_{i=1}^{N}\left[ J_1 \left(\mathbf{S}_{i,a}\cdot \mathbf{S}_{i,b} +\mathbf{S}_{i,b} \cdot \mathbf{S}_{i,c} \right)
	+ J_2 \mathbf{S}_{i,c} \cdot\mathbf{S}_{i+1,a} \right] \nonumber \\ 
	&-& H_z \sum_{j=1}^{3N} S_j^z,
\end{eqnarray}
where $\mathbf{ S}_{i,\gamma}$ is the spin-$1/2$ operator at the $\gamma$-th sublattice site of the $i$-th trimer, the intratrimer labels $\gamma \in \{a,b,c\}$ are explained
in Fig.~\ref{diagram}(a). $H_z$ represents the strength of external magnetic field, which  breaks the $\mathrm{SU}(2)$ symmetry.  The system comprises a total of $N$ trimers, resulting in a system length of $L=3N$.  The tuning parameter $g$ is defined as $g = J_2/J_1$. For simplicity, we set the
intratrimer interaction $J_1=1$ as the energy unit, so that intertrimer interaction $J_2=g$. Our interest is in the range of 
coupling ratios $g \in [0,1] $, where the system evolves between the isolated trimers and the isotropic  Heisenberg antiferromagnetic chain.

\subsection{\label{subsec:cells} Quantum phase transition}

\begin{figure*}
\includegraphics[width=16cm]{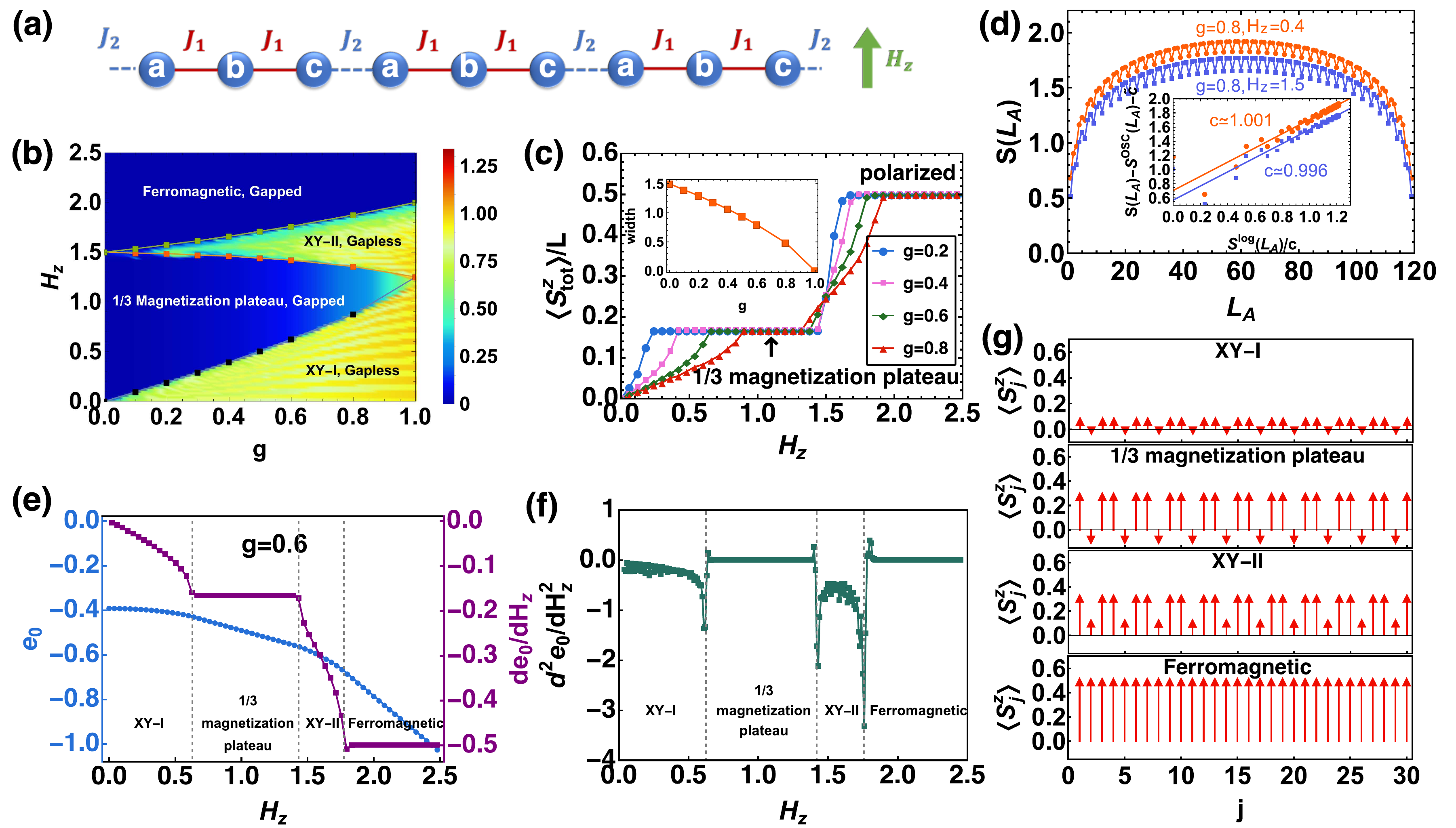}
\caption{\label{diagram} \textbf{Quantum phase transitions.} (a) Schematic representation of a trimer spin chain subjected to a longitudinal magnetic field. The analysis focuses on systems characterized by the condition $J_1 \geq J_2 >0$, with the letters $a,b,c$ denoting the three spins within a unit cell. (b) The
 phase diagram is obtained by employing the DMRG method to map the entanglement entropy onto the parameter space  $(g,H_z)$  for a system with $L=180$ spins. (c) Magnetization curves are illustrated as a function of $H_z$ for different $g$ in a system  with $L=180$. Inset shows the width of $1/3$ magnetization plateau as a function of $g$. (d) Entanglement entropy $S(L_A)$ as a function of the subsystem size $L_A$ under open boundary conditions.  Solid lines in the  inset represent the best fits to the CFT scaling form, with the optimal values for the central charges provided. (e) Ground state energy of per spin $e_0=E_0 /L$ and its first derivative $d e_0 / d H_z $ as a function of $H_z$ for the system  with $L=210$.  (f) The second derivative $d^2 e_0 / d H_z^2 $ as a function of $H_z$ for the system  with $L=210$. (g) Magnetization of each spin obtained by DMRG for four distinct phases where $g=0.6$ and $H_z =0.3, 1.0, 1.5, 2.0$. }
\end{figure*}

In the absence of a magnetic field, the antiferromagnetic quantum spin trimer chain exhibits a gapless low-energy excitation known as the two-spinon continuum \cite{cheng2022}.  When a magnetic field is applied, the $\mathrm{SU}(2)$ symmetry is broken, leading to the emergence of a quantum phase transition driven by the competition between the interaction and magnetic field.  In this subsection, we aim to investigate the detailed phase diagram using  the  DMRG method.

Quantum entanglement provides a distinctive  framework for unveiling  the ground-state properties of many-body systems and   has been extensively used  to study the quantum phase transitions \cite{RevModPhys.80.517,LAFLORENCIE20161,cheng2017,PhysRevLett.120.200602,PhysRevE.97.062134,PhysRevLett.128.020402}. Entanglement entropy, a crucial metric for assessing  bipartite quantum  entanglement, can be  easily derived from DMRG calculations. Its definition is given by
\begin{eqnarray}
	S=-\mathrm{Tr}\left[\rho_\mathrm{A} \ln \rho_\mathrm{A}\right],
\end{eqnarray}
where the reduced density matrix $\rho_\mathrm{A}$ is the partial trace of the density matrix of the whole system  $\rho$, $\rho_\mathrm{A}=Tr_\mathrm{B}\left[\rho\right]$. If $\mathrm{A}$ and $\mathrm{B}$ are entangled, the reduced density matrix must be a mixed state, and the entanglement entropy quantifies the degree of this mixing. By effectively analyzing the entanglement entropy, the characteristics of ground states in various quantum phases can be extracted. Therefore, the entanglement entropy serves as a viable and useful tool for investigating the quantum phase transitions.
As illustrated in Fig.~\ref{diagram}(b), the entanglement entropy reveals 
 four distinct phases  within the $(g,H_z)$ parameter space. When an external magnetic field is applied, a transition from the N\'{e}el phase to an incommensurate  phase  occurs,  propelling the system into the XY-I phase. However, the magnetic field is insufficient to open a gap, resulting in a ground state that remains gapless with  a nonzero entanglement entropy. In Fig.~\ref{diagram}(c), the magnetization increases  with the  magnetic field in the XY-I phase until a fractional magnetization plateau is reached.

 The fractional magnetization plateau observed in the magnetization curves can be understood through  the Oshikawa-Yamanaka-Affleck (OYA) criterion \cite{PhysRevLett.78.1984}:
\begin{equation}
	n\left(s-m\right)=\mathrm{integer}  \label{OYA}
\end{equation}
where $n$ is the number of spins in a unit cell, $m$ is the magnetization per site and $s$ denotes the magnitude of spin. For the trimer chain, it has $n=3$, $s=1/2$, so when $n(s-m)=0$  it results in $m=1/2$, which  corresponds to the full polarized state. Conversely, when $n(s-m)=1$ it  yields $m=1/6$ corresponding to the $1/3$ magnetization plateau. Fig.~\ref{diagram}(c) clearly illustrates the existence of these two plateaus. In the $1/3$ magnetization plateau phase,   the external magnetic field is insufficient to decouple the singlets.
As $g$ increases, the width of magnetization plateau  decreases,  ultimately vanishing  at $g=1$ where the trimer chain transitions into the uniform Heisenberg chain.
From the magnetization of each spin, as shown in Fig.~\ref{diagram}(g), we can observe that the magnetization  exhibits a periodic pattern corresponding to the trimerized structure. Specifically, the magnetization of the central spin in each trimer  increases as $H_z$ rises. Importantly, 
 the magnetization of  each spin remains fixed even as  the magnetic field increases  in the $1/3$ magnetization plateau phase. The ground state of an isolated trimer in the presence of a magnetic field is described by, 
\begin{equation}
\left|0\right\rangle  = \frac{1}{{\sqrt 6 }}\left( {\left| { \uparrow  \uparrow  \downarrow } \right\rangle  - 2\left| { \uparrow  \downarrow  \uparrow } \right\rangle  + \left| { \downarrow  \uparrow  \uparrow } \right\rangle } \right),
\end{equation}
which is also the  antiferromagnetic trimer state of the Haldane plateau observed in one-dimensional $ (S,s)=(1,1/2)$ mixed spin chain \cite{PhysRevB.65.214403}.  In the $1/3$ magnetization plateau phase, as shown in Fig.~\ref{diagram}(g), the expectation values of the $z$ components of three spins,  labeled $a,b,c$, are  $0.322$, $-0.144$, and $0.322$, respectively. These values are approximately coincide  with the ideal state, which has expectation values of $1/3$, $-1/6$, and $1/3$.  This  indicates that the $1/3$ magnetization plateau state exhibits  N\'{e}el order along  the direction of magnetic field, with each trimer effectively possessing a spin of $1/2$, thereby creating  the appearance of polarization for each trimer.

As the magnetic field increases, the singlets are disrupted, resulting in to the emergence of the XY-II phase.  The system remains gapless, exhibiting  a nonzero entanglement entropy in the ground state. Additionally,  the average  magnetization   is greater than that in the XY-I phase and increases with the magnetic field. As long as the magnetic field is sufficiently strong, all spins become polarized, leading to the formation of an additional magnetization plateau and the emergence of a ferromagnetic phase. We also utilize the properties of entanglement to suggest a potential conformal field theory description of the gapless phases. 
  The R\'enyi entanglement  entropy of subsystem $A$ is  defined as follows:
\begin{equation}
S_\nu \left(\rho_\mathrm{A}\right)=\frac{1}{1-\nu} \ln \left(\operatorname{Tr}\left\{\rho_\mathrm{A}^\nu\right\}\right).
\end{equation}
 In the limit $ \nu \rightarrow 1$, the above expression 
 simplifies to the von Neumann entanglement entropy,
\begin{equation}
S(\rho_\mathrm{A})=\lim _{\nu \rightarrow 1} S_\nu \left(\rho_\mathrm{A}\right)=-\operatorname{Tr}\left[\rho_\mathrm{A} \ln \rho_\mathrm{A}\right].
\end{equation}
The R\'enyi entanglement  entropy of subsystem  $A$ follows the scaling form \cite{PhysRevLett.104.095701,PhysRevB.92.054411,PhysRevB.105.014435}:
\begin{equation}
S_\nu\left(L_\mathrm{A}\right)=S_\nu^{\log }\left(L_\mathrm{A}\right)+S_\nu^{\operatorname{osc}}\left(L_\mathrm{A}\right)+\tilde{c}_\nu,
\end{equation}
where  
\begin{equation}
S_\nu^{\log }\left(L_\mathrm{A}\right)=\frac{c}{6 \eta}\left(1+\frac{1}{\nu}\right) \ln \left\{\left[\frac{\eta L}{\pi} \sin \left(\frac{\pi L_\mathrm{A}}{L}\right)\right]\right\},
\end{equation}
and 
\begin{equation}
S_\nu^{\mathrm{osc}}\left(L_\mathrm{A}\right)=F_\nu\left(\frac{L_\mathrm{A}}{L}\right)\frac{\cos \left(2 k_F L_\mathrm{A}\right)}{\left|\frac{2 \eta L}{\pi} \sin \left(\pi L_\mathrm{A} / L\right)\right|^{\frac{2 \Delta_1}{\eta \nu}}}.
\end{equation}
Here, $\eta=1,2$ corresponds to periodic and open boundary conditions, respectively. The  central charge
$c$, the Fermi momentum $k_F$, and the scaling dimension $\Delta_1$ are universal parameters. $F_{\nu} (L_\mathrm{A}/L) $ is a universal scaling function and $\tilde{c}_{\nu}$ is a nonuniversal constant. By fitting the DMRG data with these functions for $\nu =1$,   we extract the central charges of the two XY phases, which serve as indicators of their universality classes. As shown in Fig.~\ref{diagram}(d), the both  XY phases are  described by the conformal field theory with  central charges $c\simeq 1$.

Furthermore, to identify the types of quantum phase transitions present, we have conducted an analysis involving the computation of the first and second derivatives of the ground state energy with respect to the magnetic field $H_z$, see Figs.~\ref{diagram}(e) and \ref{diagram}(f). According to the Hellmann-Feynman theorem, the magnetization curves (see Fig.\ref{diagram}(c)) and the first derivative of the ground state energy $de_0 /d H_z$ exhibit the similar behaviors,
characterized by continuity but a lack of differentiability near the critical points. The second derivative $d^2 e_0 /d H_z^2$  displays   nonanalytic behavior in the vicinity of  these critical points. Collectively, these results suggest that the  quantum phase transitions between the XY-I, $1/3$ magnetization plateau, XY-II, and ferromagnetic phases are  second-order  quantum phase transitions. In Supplementary note 1, we also provide the real-space spin-spin correlation function for the four phases.  The correlation functions  in the XY-I and XY-II phases decay according to power laws, with the critical exponents converging towards $1$, which is analogous  to the $S=1/2$ isotropic Heisenberg chain \cite{PhysRevLett.76.4955}.

\begin{figure*}[t]
	\includegraphics[width=16cm]{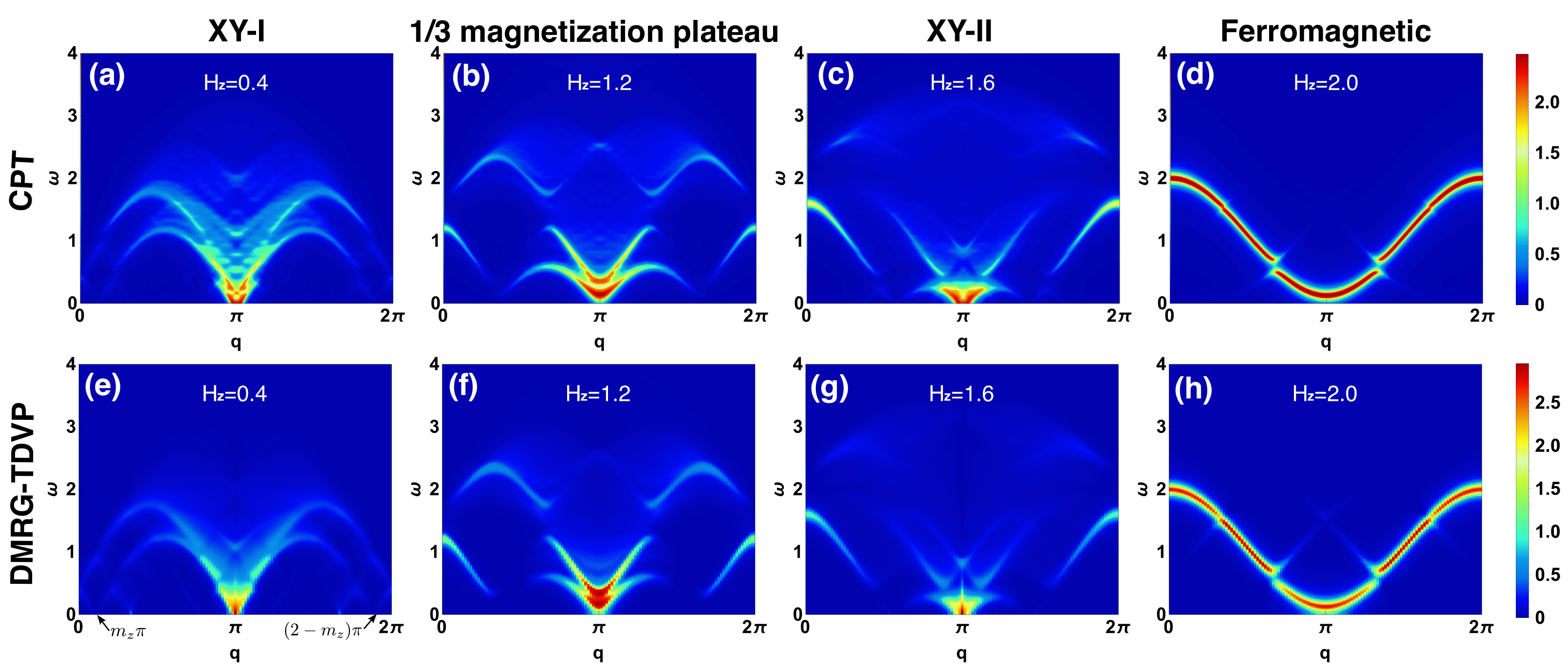}
	\caption{\label{sxx-g8} \textbf{$\mathcal{S}^{xx}  (q,\omega) $  obtained from CPT and DMRG-TDVP calculations for different phases.}  $\mathcal{S}^{xx}  (q,\omega) $ in  (a)(e) XY-I phase, (b)(f) $1/3 $ magnetization  plateau phase, (c)(g) XY-II phase, and (d)(h) Ferromagnetic phase. All results are derived  from the case where $g=0.8$, and the DMRG-TDVP calculations are  conducted for a system  with length $L=120$.  The  color coding of $\mathcal{S}^{xx} (q,\omega)$ uses a piecewise function with the boundary value $U_0=2$. 
	Below this boundary, the low-intensity portion is characterized by a linear mapping of the spectral
	function to the color bar, while above the boundary a logarithmic scale is used, $U=U_0+\log_{10}[\mathcal{S}^{zz}(q,\omega)]- \log_{10}(U_0)$. }
\end{figure*}

\subsection{Excitation spectra}

In this section, we present the spin excitation spectra of the trimer chain in an external magnetic field, utilizing the dynamical structure factor (DSF):
\begin{equation}
\mathcal{S}^{\alpha \beta}(q, \omega)=\sum_j \mathrm{e}^{-\mathrm{i} q j}\left[\int_{-\infty}^{\infty} \mathrm{d} t \mathrm{e}^{\mathrm{i} \omega t}\left\langle\hat{S}_j^\alpha(t) \hat{S}_0^\beta\right\rangle\right],
\end{equation}
where $\alpha,\beta$ refer to the spin components $x$, $y$, and $z$. We calculated the DSF using the CPT and DMRG-TDVP methods to study the spin dynamics under the modification of control parameters, $g$ and $H_z$. Detailed calculations can be found in  Sec.~\ref{method}.
In our previous study, we have utilized the quantum Monte Carlo methods with subsequent numerical analytic continuation to investigate the spin dynamics of trimer chain in the absence of magnetic field, revealing the doublons and quartons in the intermediate-energy and high-energy regimes, respectively \cite{cheng2022}.  For comparison, the spectral characteristics are also assessed using  the DMRG-TDVP and CPT calculations, with results available in Supplementary note 2. Furthermore, we provide the spin excitation spectra obtained  from  ED calculation in Supplementary note 5. 

In Fig.~\ref{sxx-g8}, the transverse excitation spectra  $\mathcal{S}^{xx}  (q,\omega) $ for four distinct phases are presented. The excitations are gapless in  two XY phases and gapped in the $1/3$ magnetic plateau and ferromagnetic phases.  To understand the spin dynamics of the XY-I and XY-II phases, we will examine the zero-energy excitations. 
The incommensurability observed in the spin dynamics of an AF spin-$1/2$ chain subjected to a longitudinal magnetic field can be interpreted using the language of spinless fermions \cite{takayoshi2023phase}. The longitudinal magnetic field acts as a chemical potential, which alters the band filling and breaks  the degeneracy of the electron-hole bands. The intraband and interband zero-energy excitations  correspond to the longitudinal and transverse fluctuations, respectively. 
For the trimer chain subjected to a longitudinal magnetic field,   incommensurability arises from the splitting of the bands. 
Figs.~\ref{sxx-g8}(a)(e) display the transverse excitations $S^{xx}(q, \omega)$  as the number of particles is varied, leading to fluctuations that reach  zero energy at incommensurate wave numbers $q=m_z \pi$ and $q=(2-m_z)\pi$ in addition to $q=\pi$. The spectral weight  is concentrated at the commensurate positions corresponding to each reciprocal lattice points at $q=\pi$ in the XY-I and XY-II phases.  In the high-energy regime, a continuum is observed  in both the $1/3$ magnetization plateau and the XY-II phases.  At the ferromagnetic phase, see
Figs.~\ref{sxx-g8}(d)(h),  all spins are polarized,  and the spin excitation continues to  propagate as   magnons.  Energy gaps are observed  at the edges of the Brillouin zone, specifically at  $q=\pi/3,2\pi/3,4\pi/3,5\pi/3 $ where spin waves are diffracted due to the periodic potential of the trimerized interaction. Consequently,  the magnons at the edges of the  Brillouin zone  exhibit two distinct energy levels for the same wave vector. In Supplementary note 3, we also present the longitudinal excitation spectrum and discuss the zero-energy excitations that correspond  to the longitudinal  fluctuations.

\begin{figure*}[t]
	\includegraphics[width=16cm]{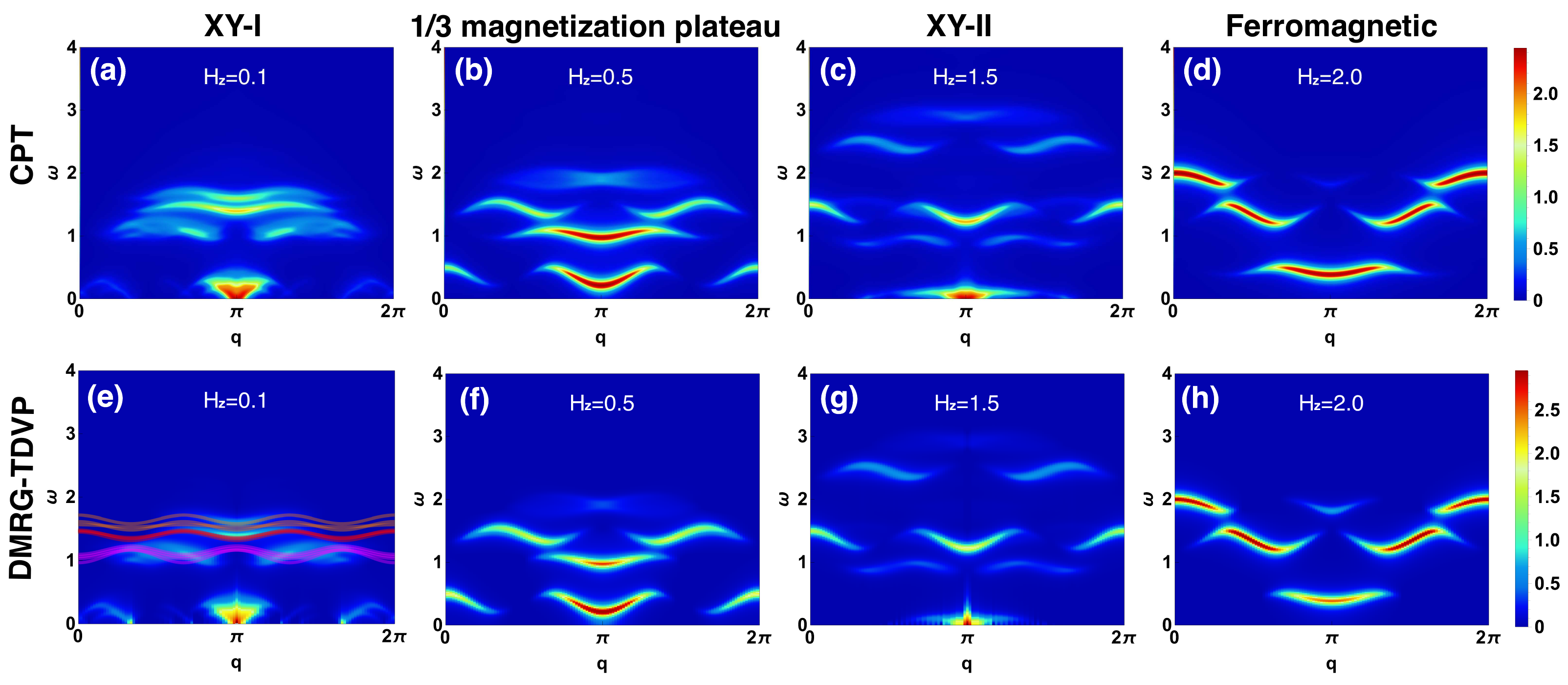}
	\caption{\label{Sxx} \textbf{ $\mathcal{S}^{xx}  (q,\omega) $  obtained from CPT and DMRG-TDVP calculations for  weak intertrimer interaction.}  $\mathcal{S}^{xx}  (q,\omega) $ in  (a)(e) XY-I phase, (b)(f) $1/3 $ magnetization  plateau phase, (c)(g) XY-II phase, and (d)(h) Ferromagnetic phase. All results are derived from the case where $g=0.3$, and the DMRG-TDVP calculations are conducted for a system with length $L=120$. The  color coding of $\mathcal{S}^{xx} (q,\omega)$ uses a piecewise function with the boundary value $U_0=2$. }
\end{figure*}

\begin{figure*}[t]
	\includegraphics[width=16cm]{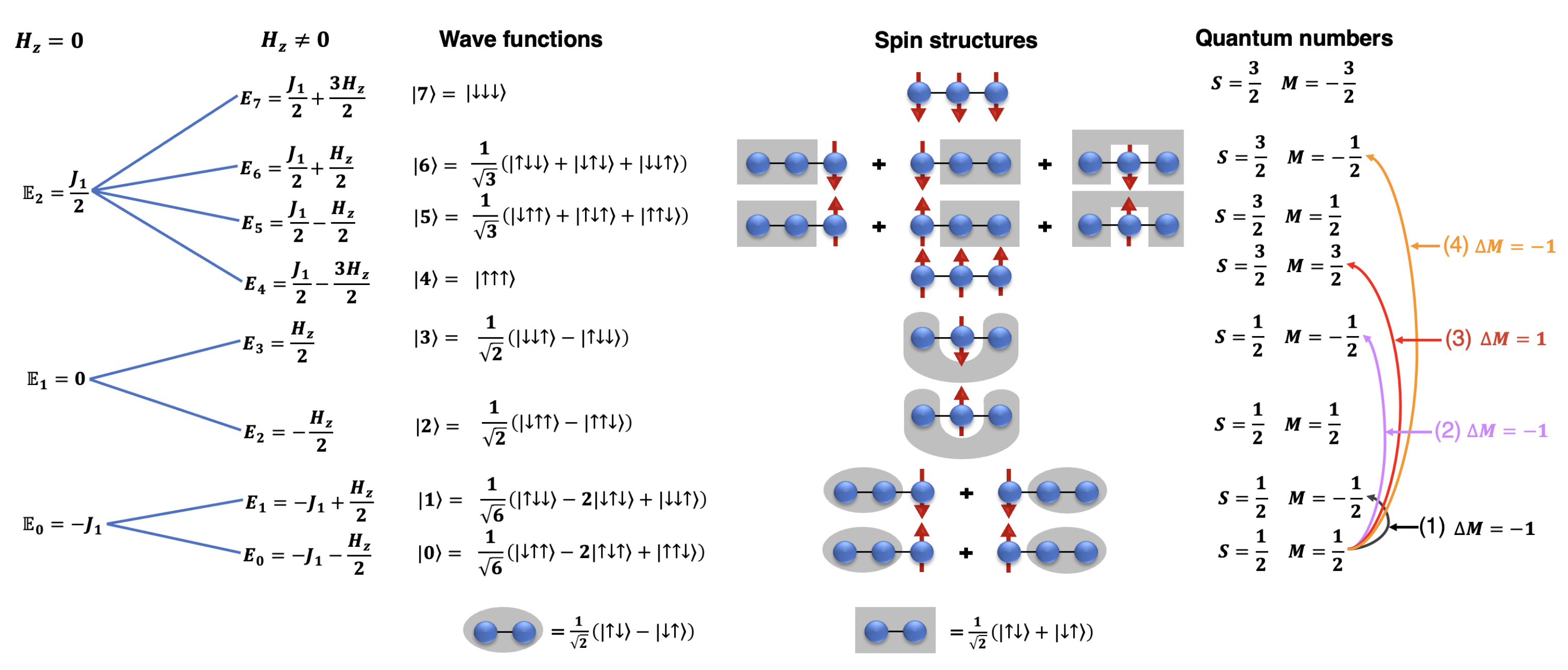}
	\caption{\label{energy} \textbf{ The level spectrum, wave functions, and quantum numbers of one isolated trimer under a longitudinal magnetic field.} The second column lists the wave functions in the spin-$z$ basis, while the third column presents the spin structures using a basis of singlets (gray ovals and rounded shapes), zero-magnetization triplets (gray square shapes), and unpaired spins (arrows). The last column lists the total spin quantum number $S$, magnetic quantum number $M$, and the internal trimer excitations with $\Delta M = \pm 1$. }
\end{figure*}

\begin{figure*}[t]
	\includegraphics[width=16cm]{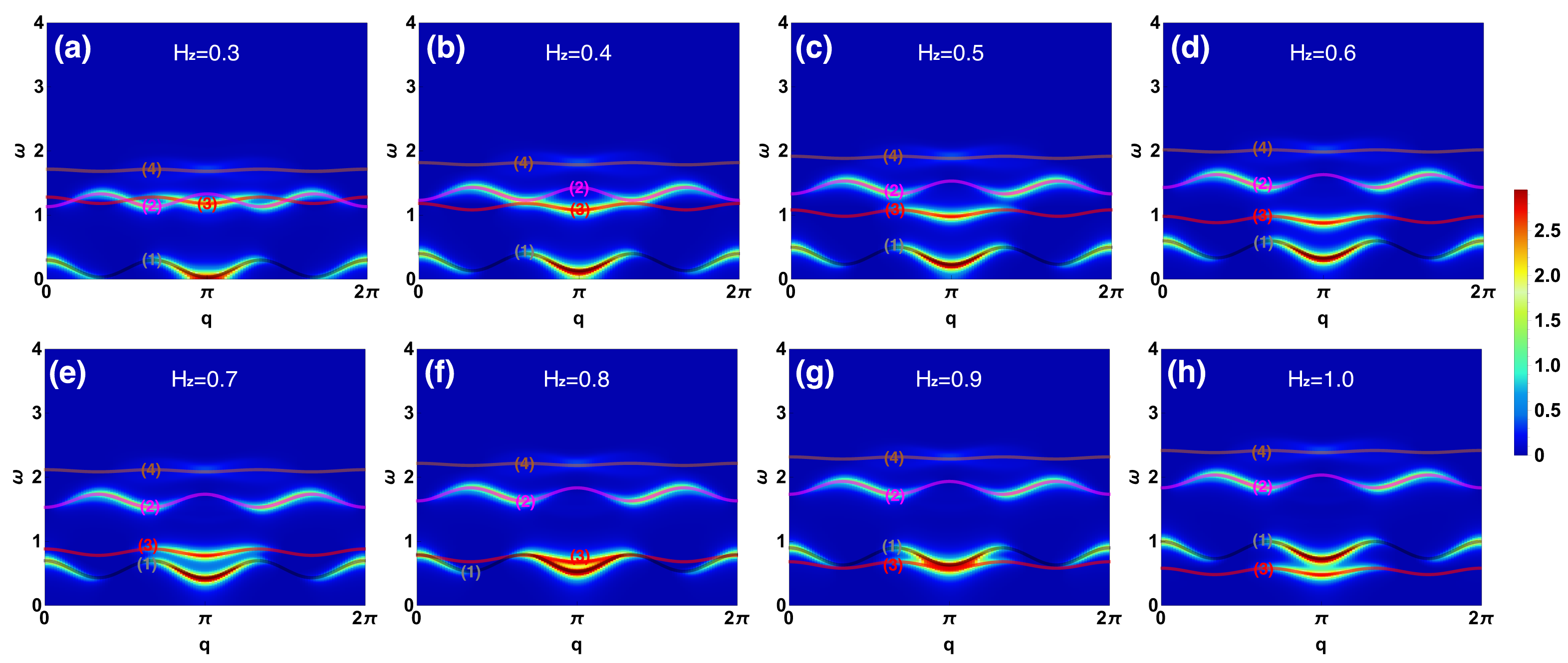}
	\caption{\label{Sxx-plateau-dispersion} \textbf{ $\mathcal{S}^{xx}  (q,\omega) $ in the  $1/3 $ magnetization  plateau phase.} All results are obtained by  DMRG-TDVP calculations  for a system with length  $L=120$, and the  color coding of $\mathcal{S}^{xx} (q,\omega)$ uses a piecewise function with a boundary value $U_0=2$. The dispersion lines, distinguished by colors and numerical labels, correspond to various localized excitations within a single trimer. (1)(2)(4) are the excitations from $\left| 0 \right\rangle \rightarrow \left| 1 \right\rangle$, $\left| 0 \right\rangle \rightarrow \left| 3 \right\rangle$ and $\left| 0 \right\rangle \rightarrow \left| 6 \right\rangle$ with $\Delta M =-1$, respectively. (3) is the  excitations from $\left| 0 \right\rangle \rightarrow \left| 4 \right\rangle$ with $\Delta M =1$. }
\end{figure*}

In our previous study \cite{cheng2022}, we found that a smaller $g$ induces  rich intermediate-energy and high-energy  excitations beyond the spin wave. Therefore, it is of great interest  to investigate the evolution of intermediate-energy  and high-energy quasiparticles, referred  to as the doublons and quartons, under the influence of a magnetic field. In this study, we focus on the weak intertrimer coupling  $g=0.3$  to examine their dynamical evolutions. 
In this case,  the system behaves as  isolated trimers, allowing for straightforward analysis. From Figs.~\ref{Sxx}(a)(e), we observe that the low-energy excitation resembles the excitation spectrum of conventional spinons in a magnetic field. A splitting of  the dispersion relation occurs, characterized by the emergent fermions of the Heisenberg chain in the presence of magnetic field \cite{PhysRevB.94.125130}. The intermediate-energy spectrum shows minimal  separation near $q=\pi/3$ and $q=5\pi/3$, with a continuum emerges, possibly due to the propagation of doublons dressed by spinons \cite{cheng2022}. The high-energy spectrum is distinctly  split  into two branches by the magnetic field.

As $H_z $ increases, see Figs.~\ref{Sxx}(b)(f), the excitation gap  opens, leading the system into a $1/3 $ magnetization  plateau phase.  Consequently, the lower spectrum of high-energy excitations with $\Delta M=1$ transitions to the low-energy regime.
When $H_z =1.5$, the system evolves into  the gapless XY-II phase (see Figs.~\ref{Sxx}(c)(g)), where the low-energy, intermediate-energy and high-energy spectra exhibit clear differentiation.  In the ferromagnetic phase $H_z=2.0$, shown in
Figs.~\ref{Sxx}(d)(h),    the spin excitation are predominantly characterized by  spin waves due to complete spin polarization.  Additionally, some energy gaps are  observed  at the   Brillouin zone  edges  $q=\pi/3,2\pi/3,4\pi/3,5\pi/3 $, which is consistent with the results of Figs.~\ref{sxx-g8}(d)(h).

The preceding discourse primarily focuses on the impact of a magnetic field on the spin excitation spectra of a trimer chain with a constant intertrimer interaction strength.  To enhance our understanding  of the spin dynamics within such a trimer chain under the influence of a magnetic field, we present additional results obtained through DMRG-TDVP calculations for varying values of the $g$ parameter. Further details are provided in Supplementary Note 4 for clarification.


\subsection{Excitations mechanisms}

To gain a deeper understanding of  the intermediate-energy and high-energy spin dynamics, it is instructive to analyze the complete level spectrum and the corresponding eigenvectors of a single trimer. As depicted in Fig.~\ref{energy}, the application of a magnetic field causes the splitting of three energy levels into eight distinct levels.  Notably,
  the eigenvectors, spin quantum numbers, and magnetic quantum numbers   remain invariant.  When $H_z \leq 1.5$, the ground state of the  trimer is denoted as $\left|0\right\rangle$ with an energy $E_0 = -J_1-H_z/2$. 
Considering the excitations with $\left|\Delta M\right|=1$ from $\left|0\right\rangle$,  only four cases satisfy this condition, as indicated  in the last column of Fig.~\ref{energy}.
For small $g$, the coupling between trimers can be treated as a perturbation of the product state of isolated trimers. This approach has been validated in our previous study of a trimer chain without an external magnetic field \cite{cheng2022}. 
Here, the perturbative analysis remains an effective tool to handle with the spin excitations of trimer chain under a magnetic field, particularly in the XY-I and $1/3 $ magnetization  plateau phases.
In the XY-I phase with a small $g$, a weak magnetic field induces an incommensurate ground state with  slight magnetization. By utilizing the  ground state $\left|0\right\rangle$ and the first excited state $\left|1\right\rangle$ of a single trimer,  we can  construct an  approximate ground state with  antiferromagnetic order, such as $\left|\psi\right\rangle_g = \left|0101\dots01\right\rangle$. Consequently,  we are able to calculate the dispersion relations corresponding to the intermediate-energy and high-energy excitations with $\left|\Delta M \right|= 1$ by employing  only $N=4$ trimers, details can be found in Supplementary note 6. 
Regarding the intermediate-energy excitations, 
four  dispersion relations 	 emerge at $\omega \propto J_1$, as shown in Fig.~\ref{Sxx}(e), which describe the localized excitations  from $\left|0\right\rangle$ to $\left|3\right\rangle$,
\begin{equation}
\label{intermediate}
	\epsilon_{\mathrm{D}} (q)= \left\{
    \begin{split}
   &-  \frac{1}{3} g \cos{(3q)}+ \mathbb{E}_1 -\mathbb{E}_0 + H_z - \frac{1}{9}g, 
    \\
    &- \frac{2}{9} g \cos{(3q)}+\mathbb{E}_1 -\mathbb{E}_0 + H_z, 
    \\
    &- \frac{2}{9} g \cos{(3q)}+\mathbb{E}_1 -\mathbb{E}_0 + H_z + \frac{2}{9}g, 
    \\
    &- \frac{2}{9} g \cos{(3q)}+\mathbb{E}_1-\mathbb{E}_0  + H_z + \frac{2}{9}g.
    \end{split}
	\right.
\end{equation}
It can be observed that these dispersion relations depend on the energy gap $\mathbb{E}_1-\mathbb{E}_0$ between the ground state and first excited-state energies of one single trimer in the absence of magnetic field,  and they increase with the application of magnetic field $H_z$.
In comparison to the case $H_z=0$ \cite{cheng2022}, only one branch of doublons remains under the influence of magnetic field; thus, the intermediate-energy excitation corresponds to the generation of doublons.
These dispersion lines do not align  well  with the spectrum due to the approximation of the ground state. Additionally,  we observe a continuum that may originate  from  bound spinons.  The central doublet,  which is dressed by these spinons,  propagates through the system, resulting in various internal modes of these composite excitations.  This, in turn, leads to a band of finite width in the energy of these excitations.  For further details,  see the propagation of doublons  in Supplementary note 2.

 For the high-energy excitations, the application of a magnetic field results in the division of the spectrum into two distinct branches,  as shown in Figs.~\ref{Sxx}(a)(e).
 Both branches originate from the high-energy internal trimer excitations. We designate the upper branch (excitation from  $\left|0\right\rangle$ to $\left|6\right\rangle$) as the  upper quarton, and the lower one (excitation from  $\left|0\right\rangle$ to $\left|4\right\rangle$) as the lower quarton.
The    dispersion relations for the  upper quarton are given by,
\begin{eqnarray}
	\label{high1}
\epsilon_{\mathrm{UQ}} (q)=\left\{
    \begin{split}
& \frac{1}{18} g \cos{(3q)}+\mathbb{E}_2 -\mathbb{E}_0 + H_z - \frac{1}{6}g,\\
& \frac{2}{9} g \cos{(3q)}+\mathbb{E}_2 -\mathbb{E}_0 + H_z - \frac{1}{6}g,\\
&\frac{2}{9} g \cos{(3q)}+\mathbb{E}_2 -\mathbb{E}_0 + H_z + \frac{1}{6}g,
\end{split}
\right.
\end{eqnarray}
and the ones for  lower quarton are given by,
\begin{eqnarray}
	\label{high2}
	\epsilon_{\mathrm{LQ}} (q)=\left\{
    \begin{split}
& \frac{1}{6} g \cos{(3q)}+\mathbb{E}_2 -\mathbb{E}_0 - H_z + \frac{1}{18}g,\\
& \frac{2}{9} g \cos{(3q)}+\mathbb{E}_2 -\mathbb{E}_0 - H_z + \frac{1}{18}g,\\
& \frac{2}{9} g \cos{(3q)}+\mathbb{E}_2 -\mathbb{E}_0 - H_z +\frac{1}{6}g.
\end{split}
\right.
\end{eqnarray}
 These dispersion relations depend on the energy gap $\mathbb{E}_2-\mathbb{E}_0$ between the ground state and second excited-state energies of one single trimer in the absence of magnetic field, and they
 exhibit a significant concordance  with the DMRG-TDVP results concerning the  positioning of these excitations and their bandwidths. This alignment indicates  the conceptualization of localized excitations is valid, despite the fact that the calculation relies on a rather coarse  approximation of the ground state.
 Consequently, the high-energy  quartons remain persist in the XY-I phase at  low values of  $g$.

In the $1/3$ magnetization plateau phase, each trimer exhibits  an effective magnetic quantum number $1/2$, resembling a polarized spin as a unit cell. We can construct the ground state of the $1/3$ magnetization plateau using the ground state of single trimer, $\left|\psi\right\rangle_g = \left|000\dots00\right\rangle$,  to study the spin dynamics.
The low-energy spin wave is generated by the flipping of one spin within the ferromagnetic state, a phenomenon effectively described  by the propagation of magnons. In this scenario, we can manipulate the effective spin of one trimer; for instance, by altering  one trimer from $\left|0\right\rangle$ to $\left|1\right\rangle$ in state $\left|\psi\right\rangle_g $, that results in the dispersion relation, 
\begin{eqnarray}
	\epsilon^{(1)}(q)=	\frac{4}{9} g \cos{(3q)}+ H_z - \frac{4}{9}g,
\end{eqnarray}
which coincides well with the low-energy excitation spectrum, irrespective of the magnitude of the magnetic field, as illustrated in  Fig.~\ref{Sxx-plateau-dispersion}. We designate this excitation as the reduced spin wave  inspire of the conventional magnon picture. Moving on to the intermediate-energy excitations, where one trimer is excited from $\left|0\right\rangle$ to $\left|3\right\rangle$ with $\Delta M=1$, the associated  dispersion relation is 
\begin{eqnarray}
	&&\epsilon^{(2)}(q)=	-\frac{1}{3} g \cos{(3q)}+ \mathbb{E}_1-\mathbb{E}_0+H_z-\frac{2}{9}g.
\end{eqnarray}
Here, the intermediate-energy mode is termed as the doublon rather than the magnon, as it arises from the excitation of localized trimers and exhibits a higher energy gap compared to the low-energy magnon. 
For the high-energy excitations, two distinct branches of the excitation spectra  emerge, which  corresponds to the  excitation $\left|0\right\rangle \rightarrow \left|6\right\rangle$ with $\Delta M=-1$ and $\left|0\right\rangle \rightarrow \left|4\right\rangle$ with $\Delta M=1$. The corresponding  dispersion relations are given by,
\begin{eqnarray}
	&&\epsilon^{(3)}(q)=\frac{1}{6} g \cos{(3q)}+ \mathbb{E}_2-\mathbb{E}_0-H_z+\frac{1}{9}g,\\
	&&\epsilon^{(4)}(q)=\frac{1}{18} g \cos{(3q)} + \mathbb{E}_2 -\mathbb{E}_0+H_z-\frac{1}{3}g,
\end{eqnarray}
which are referred to as  the high-energy  quartons.   Notably, 
it can be observed that  the reduced spin wave ($\left| 0 \right\rangle \rightarrow \left| 1 \right\rangle$), doublon ( $\left| 0 \right\rangle \rightarrow \left| 3 \right\rangle$) and upper quarton ( $\left| 0 \right\rangle \rightarrow \left| 6 \right\rangle$) share the same magnetization quantum number $\Delta M=-1$, and they collectively increase in energy as the magnetic field intensifies. Conversely,  the lower quarton  descends independently  due to its distinct magnetization quantum number $\Delta M=1$, ultimately becoming the low-energy spectrum when $H_z \geq 0.9$. More interestingly, as shown in Figs.~\ref{Sxx}(c)(g), it is noteworthy  that even in the XY-II phase, the excitations characterized by $\Delta M=-1$  remain observable in the high-energy regime.

\begin{figure*}[t]
	\includegraphics[width=16cm]{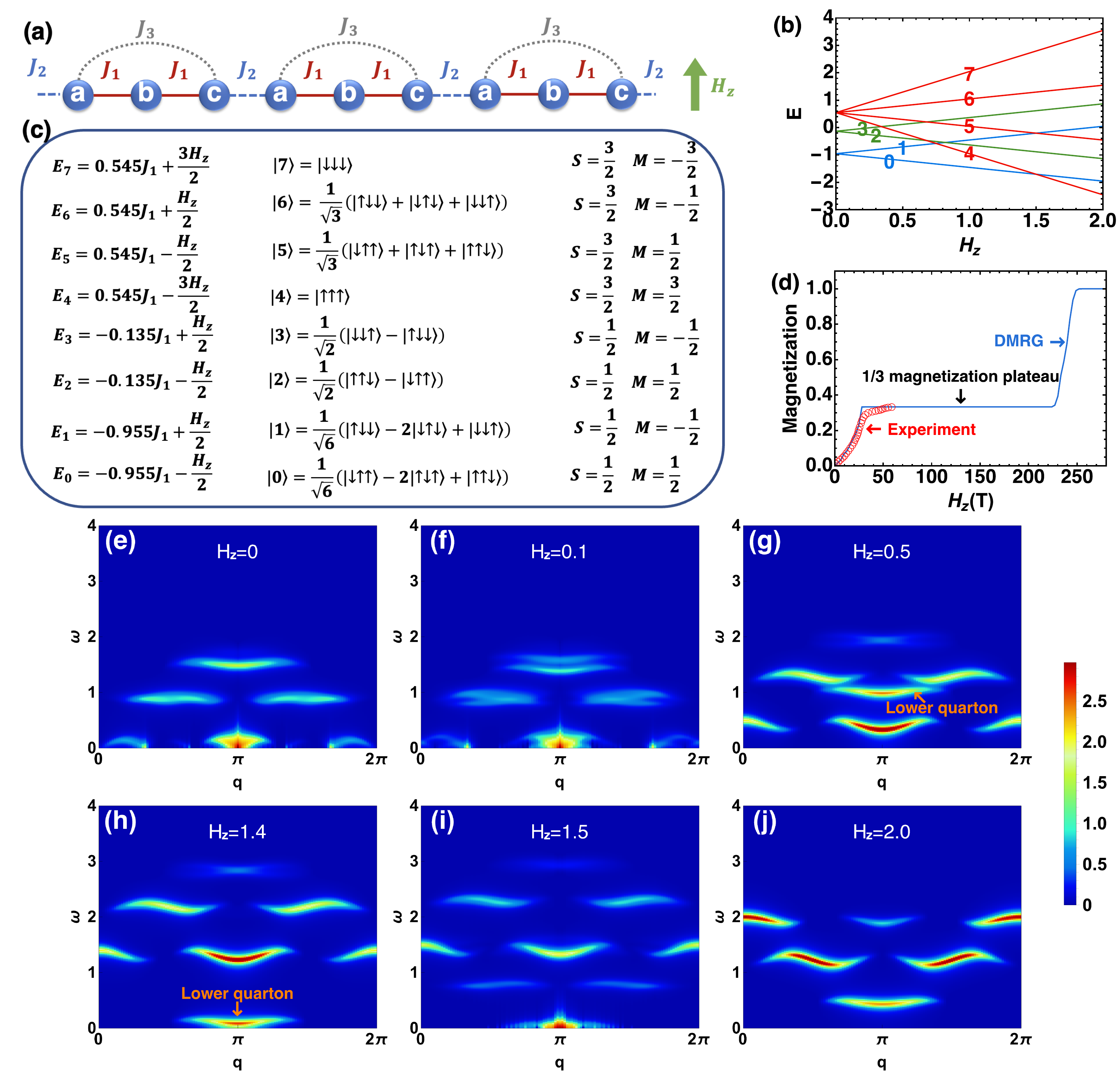}
	\caption{\textbf{Trimer model related to the experimental material \ce{Na_2Cu_3Ge_4O_12}.} (a) Schematic representation of the trimer model with next-nearest neighbor intratrimer exchange couplings $J_3$, and $J_2=J_3=0.18J_1$. The spins labeled as $a$, $b$ and $c$ are the three \ce{Cu^2+} spins within a trimer unit. (b) Energy levels as functions of the magnetic field $H_z$.
	(c) Eigenevergies, wave functions, and quantum numbers of an isolated trimer unit in the material under the magnetic field $H_z$. (d) The magnetization curves obtained by experimental measurements and DMRG calculation, with the results normalized to the maximum of magnetization. The experimental data is extracted from Fig.1d in Ref.\cite{bera2022emergent}.
	$\mathcal{S}^{xx}  (q,\omega) $ of spin model related to experimental materials \ce{Na_2Cu_3Ge_4O_12 } in  (e) the case without magnetic field, (f) XY-I phase, (g)(h) $1/3 $ magnetization  plateau phase, (i) XY-II phase, and (j) Ferromagnetic phase obtained by  DMRG-TDVP calculation for $L=120$. The  color coding of $\mathcal{S}^{xx} (q,\omega)$ uses a piecewise function with  a boundary value $U_0=2$.}
	\label{material}
\end{figure*}

\subsection{Quantum  magnets}

It has been  discovered that  \ce{Na_2Cu_3Ge_4O_12 } is an ideal realization of the spin-$1/2$ antiferromagnetic trimer chain, wherein  \ce{Cu_3O_8 } comprises  the  trimers formed by three edge-sharing \ce{CuO4} square planes  arranged linearly. The magnetic \ce{Cu^2+} ions within
the \ce{CuO4} square planes  demonstrate quantum spin-$1/2$ characteristics \cite{JAP2014,bera2022emergent}.  Fig.~\ref{material}(a)  presents  a more realistic spin model that incorporates  an additional next-nearest neighbor intratrimer exchange coupling $J_3$. The Hamiltonian for this system is given by 
\begin{eqnarray}
	\mathcal{H}^\prime&=&\sum_{i=1}^{N} [ J_1 \left(\mathbf{S}_{i,a}\cdot \mathbf{S}_{i,b} +\mathbf{S}_{i,b} \cdot \mathbf{S}_{i,c} \right)
	+ J_2 \mathbf{S}_{i,c} \cdot\mathbf{S}_{i+1,a} \nonumber \\
	 &+& J_3 \mathbf{S}_{i,a} \cdot\mathbf{S}_{i,c}]
	-H_z \sum_{j=1}^{3N} S_j^z,
\end{eqnarray}
where the experimental measurements have established the coupling strengths as $J_1 =235K$ and $J_2=J_3=0.18J_1$. When $H_z =0$, the eight energy levels (see Fig.~\ref{material}(c)) reduce to  three ones, $E_0^{\prime}=-0.955J_1$, $E_1^{\prime}=-0.135J_1$ and $E_2^{\prime}=0.545J_1$. The doublon and quarton  manifest at $\omega \sim E_1^{\prime}-E_0^{\prime}=0.82J_1 $  and $\omega \sim E_2^{\prime}-E_0^{\prime}=1.5J_1 $, respectively, as shown in Fig.~\ref{material}(e).   In the  inelastic neutron scattering measurements conducted  on \ce{Na_2Cu_3Ge_4O_12} \cite{bera2022emergent},  three excitation modes have been identified. The two-spinon modes are observed below $5$ meV, while,  the doublon and quarton states appear in the intermediate ($17-22$ meV) and high ($32-37$ meV) energy ranges, respectively. These findings are corroborated  by the intermediate-energy (at $\omega \sim 0.82J_1$) and high-energy (at  $\omega \sim 1.5 J_1$) excitations observed in our numerical simulations.

In Fig.~\ref{material}(b), the application of a magnetic field results in the splitting of the three energy levels of a single trimer into eight distinct levels. When $H_z \leq 1.5$, 
the ground state is $\left|0\right\rangle$ with an energy of $E_0 =-0.955 J_1 -H_z /2$. 
 Although the antiferromagnetic interaction $J_3$ competes with the interaction $J_1$ and induces  frustration within the spin system, when $J_3 \leq J_1$, introducing $J_3$ does not change the relative sequence of energy levels and their quantum numbers. There are only minor shifts in their eigenvalues,
 as depicted in  Fig.~\ref{material}(c). Thus, the spin excitations can still be characterized by the  quasiparticles doublons and quartons.
The trimer chain subjected to the $J_3$ interaction continues to  display a $1/3$ magnetization plateau, as shown in  Fig.~\ref{material}(d).  Experimental measurements have confirmed the presence of $1/3$ magnetization plateau above 28 Tesla \cite{bera2022emergent}, which aligns with our DMRG calculations. Although
 the phase diagram and $1/3$ magnetization plateau have been elucidated \cite{bera2022emergent}, significantly less is   understood regarding  the  evolution of intermediate-energy and high-energy excitations   under the influence of a magnetic field. In this subsection, we present the  excitation spectrum  of the model depicted in Fig.~\ref{material}(a), which is pertinent to the material \ce{Na_2Cu_3Ge_4O_12}.  Due to the weak $J_3$,  the spin excitations  in four phases displayed in Figs.~\ref{material}(f)-(j) exhibit similarities to those of a  trimer chain devoid of $J_3$. This includes  the separation of high-energy spectra, the presence of gapless excitations in the XY-I and XY-II phases, and the emergence of a gap at the edges of Brillouin zones in the Ferromagnetic phase. Our theoretical results concerning the high-energy quasiparticles excitations under the magnetic field can be directly  validated through the inelastic neutron scattering experiments on the material \ce{Na_2Cu_3Ge_4O_12}.

 The realization of doublon and quarton in specific experimental materials drives us to think deeper into how we might manipulate these excitations and explore their possible applications. Among that, the BEC of doublon or quarton is quite an important topic.  BEC represents a compelling  state of matter that has been observed in bosonic atoms and cold gases. Quasiparticles associated with  magnetic excitations, which  possess integer spin and adhere to Bose statistics, such as the magnon and triplon, are integral to the study of   BEC \cite{ruegg2003bose,giamarchi2008bose,zapf2014bose,matsumoto2024quantum}. In particular, in the dimerized antiferromgnets like \ce{TlCuCl_3}, the  intradimer interaction is stronger than the interdimer interaction, thereby an isolated dimer exhibits a singlet ground state characterized by a total spin $S=0$ and a triplet excited state with spin $S=1$. Due to the relatively weak interdimer interaction, the magnetic excitations are predominantly governed  by triplons. After applying a magnetic field, the Zeeman term controls the density of triplons, resulting in a decrease in energy for  the triplon with a magnetic quantum number $S^z=1$. At a critical magnetic field $H_{C1}$, the energy of the triplons reaches zero,  leading to their gradual condensation into  the ground state until a second critical magnetic field $H_{C2}$ is attained. Beyond $H_{C2}$, all spins become polarized. 
 
 Inspired by the triplon BEC phenomenon, it is pertinent to inquire that  whether quarton BEC can be observed in trimerized systems. In this context, we present a preliminary  analysis of  quarton BEC based on our findings.  Firstly, the high-energy quartons arise from the internal trimer excitations and possess an integer spin quantum number $S=1$,  thereby conforming the Bose statistics. Secondly,  the application of a  magnetic field causes    the lower branch of quartons to approach zero energy, as show in Fig.~\ref{Sxx-plateau-dispersion} and Figs.~\ref{material}(g)(h). At the critical point $H_{C1}$, which delineates the transition between  the $1/3$ magnetization plateau phase and the XY-II phase, the lower quartons  commence condensation and accumulation as the magnetic field  intensifies within  the XY-II phase.

 Furthermore, although the BEC is more possible to be observed in the real materials exhibiting  3D spin systems,  the 1D and 2D limits can serve as effective  starting points for comprehending the field-induced  quasiparticle
BEC \cite{zapf2014bose,volkov2020magnon,matsumoto2024quantum,PhysRevB.107.205150}. In the context of 1D systems, it is well-established that BEC does not occur due to the significant quantum fluctuations. For 2D systems,
the presence  of a finite density of states at zero energy poses a barrier to  form the BEC. Nevertheless, the critical exponents associated with the 1D or 2D quantum critical points can be observed over a substantial temperature range \cite{zapf2014bose}. In the  2D quantum dimer magnets, both the critical field  and the critical temperature of the BEC dome can be accurately characterized \cite{PhysRevB.107.205150}. Consequently,  the studies of 1D and 2D systems are instrumental in the exploration  of quasiparticle BEC. Our recent investigation into  spin dynamics in 2D trimer systems has revealed the emergence of  high-energy quartons \cite{chang2023magnon}.  These 2D trimer systems offer valuable platforms for further examination of the quarton BEC. 
Moreover, a small but finite interlayer coupling in a quasi-2D magnet  stabilizes marginal BEC \cite{matsumoto2024quantum}.
Therefore,  a field-induced quarton BEC may emerge in the quantum material  \ce{Na_2Cu_3Ge_4O_12}. However, a technical challenge arises due to the fact that the critical magnetic field $H_{C1}$ (see Fig.~\ref{material}(d)) may exceed the limits of experimental accessibility. Therefore, only  the transition between the XY-I phase and 
the $1/3$ magnetization plateau phase  can be realized in   \ce{Na_2Cu_3Ge_4O_12} through the currently available  magnetic field.   A quantum material characterized by trimer structures and relatively weak intratrimer interactions is essential to  experimentally investigate the  quarton BEC. Theoretically, it is of interest to seek further evidence of quarton BEC by examining the $(T,H_z)$ phase diagram and the power-law temperature dependence of thermodynamic properties in both 2D and 3D trimer systems in  future studies.

\section{Discussion}\label{discussion}
In summary, we have utilized the ED, CPT and DMRG-TDVP methods to investigate the quantum phase transition and spin dynamics of the antiferromagnetic trimer chain subjected to a  longitudinal magnetic field. 
Our findings reveal that the interplay of magnetic field and interaction leads to the emergence of four distinct phases: XY-I, $1/3$ magnetization plateau, XY-II, and ferromagnetic phases.  By mapping the   entanglement entropy onto the parameter space $(g,H_z)$, we obtain a comprehensive phase diagram. Furthermore, it has been confirmed that the critical phases XY-I and XY-II phases are both characterized by the conformal field theory with a central charge $c \simeq 1$. The  transitions between these phases are identified as the second-order quantum phase transitions. 

 In the context of transverse and longitudinal excitations of a trimer chain subjected to a magnetic field, we have determined  that the incommensurate wave numbers corresponding to zero energy are dependent of the magnetization in the XY-I and XY-II phases.  In addition, we have observed the presence of gapped excitations within both the $1/3$ magnetization plateau and ferromagnetic phases. Specifically, a continuum of excitations is observed in  the high-energy regime of the $1/3$ magnetization plateau phase. In  the ferromagnetic phases, the excitations continue to be characterized   by spin waves, however,  magnons located at the  edges of the Brillouin zone   exhibit two diverse energies for the same wave vector.

 Furthermore, we have identified   the intermediate-energy and high-energy excitations for small $g$, and have elucidated their excitation mechanisms    within the XY-I and $1/3$ magnetization plateau  phases by analyzing  their dispersion relations. 
 In these phases,  the intermediate-energy and high-energy modes correspond to the propagating internal trimer excitations, referred to as doublons and quartons, respectively.
 In comparison to the trimer chain in the absence of a magnetic field, the high-energy  spectrum exhibits a splitting two branches, which correspond  to the upper quarton and lower quarton, respectively. As the magnetic field increases,  the gap between these two branches widens, resulting in the lower quarton becoming the low-energy spectrum.

 Experimentally, there are existing examples of coupled-trimer quantum magnets, such as   \ce{A_3Cu_3(PO_4)_4}
 (\ce{A=Ca, Sr, Pb}) \cite{PhysRevB.71.144411,drillon19931d,belik2005long,PhysRevB.76.014409} and \ce{Na_2Cu_3Ge_4O_12 } \cite{bera2022emergent}. Although  the trimers in  \ce{Pb_3Cu_3(PO_4)_4} do not exhibit a linear arrangement,
 two flat excitations at approximately $ \omega \sim 9 meV$ and $ \omega  \sim 13.5 meV$  have been observed in the inelastic neutron-scattering spectra measured at $8K$   \cite{PhysRevB.71.144411}. These excitations  are closely associated with the intermediate-energy (at $ \omega \sim J_1$) and the high-energy (at $ \omega \sim 1.5 J_1$) excitations  in the trimer chain without magnetic field \cite{cheng2022}. Additionally, for the quantum magnet \ce{Na_2Cu_3Ge_4O_12},  an additional next-nearest neighbor interaction  is present within the trimers; however,  its strength is relatively weak, and  the wave functions  and quantum numbers of a single trimer remain invariant.  Consequently, the doublons and quartons have been observed in the inelastic neutron-scattering experiments \cite{bera2022emergent}. We have theoretically demonstrated that  doublons and quartons remain observable in the trimer chain under a magnetic field, even with the introduction of the interaction $J_3$.  
 These findings can be directly investigated through the inelastic neutron-scattering experiments  conducted on the aforementioned  quantum materials. 
 Moreover, based on the results obtained from the 1D trimer chain under a magnetic field, it is  probable that the quarton BEC may be  identified in  experiments once suitable  materials are found.
 Our results will be  instrumental in interpreting inelastic neutron scattering and other experiments that
 probe the high-energy excitations beyond spin waves and spinons, as well as in facilitating detailed investigations of coexisting exotic excitations.

\section{Methods}\label{method}
\subsection{Matrix product states}

DSF is a significant physical quantity for studying spin dynamics, a process that been effectively facilitated by the DMRG in conjunction with the time evolution algorithms \cite{PhysRevLett.93.076401,PhysRevB.79.245101,PhysRevB.94.085136,PAECKEL2019167998,PhysRevLett.125.187201,PhysRevB.108.L220401}.
In this article, we primarily employ the TDVP method to  handle with  the time evolution of many-body systems \cite{PhysRevLett.107.070601, PhysRevB.94.165116}. Specifically, we conduct DMRG-TDVP calculations on a finite chain with open boundary conditions to analyze the spectrum. We denote the ground state of the trimer chain in the presence of a magnetic field as $\left| \mathcal{G} \right\rangle$, which allows us to compute the real time evolution of the correlation function,
\begin{equation}
\left\langle \mathcal{ G} \left|S_j^{\alpha}(t) S_0^{\beta}(0)\right| \mathcal{G} \right\rangle=\mathrm{e}^{\mathrm{i} E_0 t}\left\langle \mathcal{G}\left|S_j^{\alpha} \mathrm{e}^{-\mathrm{i} \mathcal{H} t} S_0^{\beta}\right| \mathcal{G}\right\rangle,
\end{equation}
for various times $t$ and distances $j$.  $E_0$ represents the ground state energy. We  select the site at the center of the chain designated as  the site  index $0$.  Firstly, we obtain the ground state $\left| \mathcal{G} \right\rangle$ by employing the DMRG method. Subsequently, we   introduce a local perturbation $\hat{S}_0^\beta$ at the center of  the spin chain to generate the initial state
 \begin{equation}
	\left| \phi \right\rangle = \hat{S}_0^\beta \left|  \mathcal{G} \right\rangle
 \end{equation}
 for real-time evolution. The real-time evolution state 
  \begin{equation}
	\left| \phi(t) \right\rangle = \mathrm{e}^{-\mathrm{i} \mathcal{H} t} \left| \phi \right\rangle
  \end{equation}
 is carried out using the single-site TDVP with a time step of $dt=0.05J_1^{-1}$ and a maximum time $t_{\rm{max}}=200J_1^{-1}$. Finally, a Fourier transformation is performed to obtain $\mathcal{S}^{\alpha \beta} (q, \omega)$
\begin{equation}
\mathcal{S}^{\alpha \beta}(q, \omega)=\sum_j \mathrm{e}^{-\mathrm{i} q j}\left[\int_{-\infty}^{\infty} \mathrm{d} t \mathrm{e}^{\mathrm{i} \omega t}\left\langle\hat{S}_j^\alpha(t) \hat{S}_0^\beta\right\rangle\right].
\end{equation}

Technically, to mitigate the limitations imposed by the finite-time limit constraint  on the resolution of the spectral functions in frequency space,  a Gaussian windowing function,  expressed as $\mathrm{exp}\left[-4(t/t_{max})^2 \right]$,  is incorporated  in the reconstruction of the DSF \cite{PhysRevLett.93.076401}. During the DMRG calculations, we set $\varepsilon_{\rm{SVD}}=10^{-11}$ and retained  a maximum of  $6000$ states. The time evolution is executed on a chain with open boundary conditions and $N=120$ spins, which is  sufficiently large to mitigate finite-size effects, with the maximum bond dimension  set to $2000$. All MPS simulations are performed using the ITensor library \cite{SciPostPhysCodeb.4}. 
 
 \subsection{ Cluster perturbation theory}
Cluster perturbation theory (CPT) is a theoretical framework used to study the electronic and magnetic properties of strongly correlated electrons~\cite{PhysRevB.48.418,PhysRevLett.84.522,RevModPhys.77.1027,PhysRevB.98.134410}, especially for calculating the single-particle spectral functions of Hubbard-type fermionic models and the dynamical spin structure factors of Heisenberg models. The fundamental concept of  CPT  involves partitioning a large system into smaller clusters, accurately calculating   the properties of these clusters, and then use the mean-field and perturbation theory to infer the properties of the entire system. Here, we employ ED as a computational method to determine the dynamical spin structure factor within the cluster framework. Following Ref.~\onlinecite{PhysRevB.98.134410}, we provide a concise overview of the procedural  steps involved in applying cluster perturbation theory to  spin models.

Firstly, we transform the spin model into a hard-core boson model  through the application of the following mapping,
\begin{eqnarray}
	S_i^+=b^\dagger, S_i^-=b, S_i^z=b_i^{\dagger}b_i-1/2,
\end{eqnarray}
Then the Hamiltonian can be rewrited as,
\begin{eqnarray}
	\mathcal{H}&=&\sum_{i=1}^{N}  \left( \frac{J_1}{2} b_{i,a}^\dagger b_{i,b} + \frac{J_1}{2} b_{i,b}^\dagger b_{i,c} + \frac{J_2}{2} b_{i,c}^\dagger b_{i+1,a} + H.c. \right) \nonumber \\
    &+&\sum_{i=1}^{N}\left[ J_1 n_{i,a} n_{i,b}+J_1 n_{i,b} n_{i,c}
	+ J_2 n_{i,c}n_{i+1,a} \right]  \nonumber \\
 &-& \left(H_z + \frac{J_1 + J_2}{2}\right) \sum_{i=1,i\in a/c}^{N} n_i \nonumber \\
	&-& \left(H_z + J_1\right) \sum_{i=1,i\in b}^{N} n_i + \mathcal{H}_\mathrm{const.},
\end{eqnarray}
where the $b_i^\dagger$, $b_i$ and $n_i=b_i^\dagger b_i$ are the bosonic operators with hard-core constraint $n_i=0$ or $1$.

Secondly, we split the system into clusters. In our calculations, the cluster size is chosen to be $N=8$, $ L=24$ which is sufficiently large  to  yield the accurate results. For the interaction bonds connecting adjacent clusters. We use self-consistent mean-field treatment to decouple the interactions between the clusters,
\begin{eqnarray}
	J_2 n_{1,c}n_{N,a}\approx J_2\left(\left\langle n_{1,a}\right\rangle n_{N,c}+\left\langle n_{N,c}\right\rangle n_{1,a}\right).
\end{eqnarray}

Thirdly, we employ exact diagonalization to self-consistently obtain the mean-field potentials of the two end sites, $\left\langle n_{1,c}\right\rangle$ and $\left\langle n_{N,c}\right\rangle$. And after the convergence, we run a ED simulation to obtain the real-frequency single-particle Green function matrix $\mathbf{G}_{ij}^{C}(\omega)$ using the Lanczos iteration method, where $C$ denotes the Green function matrix of the cluster.

Fourthly, the original lattice Green function matrix can be derived from the cluster Green function matrix by disregarding the nonlocal self-energy contributions between clusters.
\begin{eqnarray}
	\mathbf{G}^{L,-1}(\tilde{q},\omega)=\mathbf{G}^{C,-1}(\omega)-V(\tilde{q}).
\end{eqnarray}
where $L$ denotes the Green function matrix of the original lattice, $\tilde{\mathbf{q}}$ is the wave vector within the Brillouin zone of the supercell formed by the cluster.

Fifthly, we perform the reperiodization of the Green function matrix to restore  translational invariance. 
\begin{eqnarray}
	G_\mathrm{CPT}(q, \omega)=\frac{1}{N_s}\sum_{ij} \mathrm{e}^{-\mathrm{i}\mathbf{q}(\mathbf{r}_i-\mathbf{r}_j)}G_{ij}^L(\omega).
\end{eqnarray}
Then the transverse dynamical spin structure factor can be obtained via
\begin{eqnarray}
	S^{+-}(q, \omega)=-\frac{1}{\pi} \mathrm{Im} G_\mathrm{CPT}(q, \omega).
\end{eqnarray}

The cluster perturbation theory applied to spin models has been successfully applied to investigate the $J_1-J_2$ and $J_1-J_3$ models on a 2D square lattice~\cite{PhysRevB.98.134410,PhysRevB.106.125129}, the $J_1-J_2$ model on honeycomb lattice~\cite{PhysRevB.105.174403}, as well as 2D trimer models~\cite{chang2023magnon}. This methodology demonstrates efficacy in characterizing the continua present  in quantum spin liquid phases, as well as the magnon and triplon excitations observed in conventional N\'{e}el and valence bond solid phases. The method is exact in two limiting  scenarios: one in which  the interactions between clusters approach zero or are exceedingly weak, and the other in witch  cluster size approaches infinity or is significantly large. In the context of our trimer chain model, when $g=J_2/J_1$ is small, accurate results can be achieved even with elatively small clusters, such as $N=2,4$. The  selection of a cluster size of $N=8$ in our study ensures accuracy across both small and large $g$ regimes. This cluster size is sufficiently large to guarantee precision across a wide range of parameter values.

\section{ Data Availability}
The data that support the findings of this study are available from the corresponding authors upon reasonable request.

\section{Code availability}
The code used for the analysis is available from the authors upon reasonable request.

\begin{acknowledgments}
We like to thank Zijian Xiong for fruitful discussions. 
This project is supported by the National Key R$\&$D Program of China, Grants No. 2022YFA1402802, No. 2018YFA0306001, NSFC-92165204, NSFC-11974432, and Shenzhen Institute for Quantum Science and Engineering (Grant No. SIQSE202102), Guangdong Provincial Key Laboratory of Magnetoelectric
Physics and Devices (No. 2022B1212010008), and  Guangdong Fundamental Research Center for Magnetoelectric Physics. J.Q.C. is supported by the National Natural Science Foundation of China through Grants No. 12047562.  H.Q.W. is supported by the National Natural Science Foundation of China through Grants No. 12474248, GuangDong Basic and Applied Basic Research Foundation (No. 2023B1515120013) and Youth S$\&$T Talent Support Programme of Guangdong Provincial Association for Science and Technology (GDSTA) (No. SKXRC202404).    J.Q.C. also acknowledges the financial support from the Special Project in Key Areas for Universities in Guangdong Province (No. 2023ZDZX3054) and the Dongguan Key Laboratory of Artificial Intelligence Design for Advanced Materials (DKL-AIDAM).
\end{acknowledgments}

\section{ AUTHOR CONTRIBUTIONS}
 D.X.Y., H.Q.W., and J.Q.C. conceived and designed the project. J.Q.C., H.Q.W. and Z.Y.N. performed the numerical simulations. D.X.Y., J.Q.C., H.Q.W., and Z.Y.N.  provided the explanation of the numerical results. All authors contributed to the discussion of the results and wrote the paper.

\section{ Competing Interests}
The authors declare no competing interests.

\section{References}


\newpage
	
\begin{center}
	{\centering \bf Supplementary Note 1: Spin-spin correlation function of ground state}
	\end{center}
  
  In the main text, we present the magnetization curves, entanglement entropy, and central charge as evidence supporting the existence of four distinct phases: XY-I, $1/3 $ magnetization  plateau, XY-II, and ferromagnetic phases. To characterize these phases, we utilize the spin-spin correlation function of the ground state, denoted as $\langle S_0^x S_j^x\rangle $, as illustrated in the Supplementary Fig.~\ref{Sup-fig1}. In the XY-I and XY-II phases, magnetic ordering is absent in the ground state, and the spin correlation functions exhibit  decay according to the power laws $\langle S_0^x S_j^x\rangle \sim j^{-\eta}$ with $\eta_1 \simeq 0.995$ and $\eta_2 \simeq 0.937$. Both of these critical exponents  approach the value of  $1$,  which is consistent with the behavior observed in the $S=1/2$ isotropic Heisenberg chain \cite{PhysRevLett.76.4955}. In the $1/3 $ magnetization  plateau phase, the correlation function  $\langle S_0^x S_j^x\rangle $ rapidly  decays to zero as a function of distance, whereas in the ferromagnetic phase, $\langle S_0^x S_j^x\rangle $ remains zero for all distances. Therefore, the spin-spin correlation function $\langle S_0^x S_j^x\rangle $  serves as a valuable tool for distinguishing between the magnetically ordered phases and critical phases within this trimer spin chain subjected to a magnetic field.

	\begin{figure*}[t]
	  \includegraphics[width=18cm]{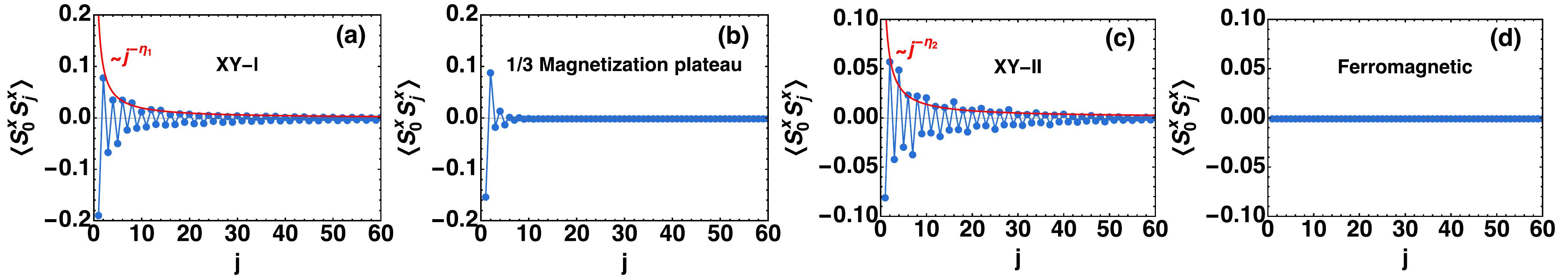}
	  \caption{\label{Sup-fig1} \textbf{Spin-spin correlation function of ground state $\langle S_0^x S_j^x\rangle $  obtained from DMRG calculations.}  $\langle S_0^x S_j^x\rangle $ as functions of distances between spins for   (a) XY-I phase ($H_z =0.2$), (b) $1/3 $ magnetization  plateau phase ($H_z =1.0$), (c) XY-II phase ($H_z =1.5$), and (d) Ferromagnetic phase ($H_z =2.0$). The red lines are the fits of correlation functions, $\langle S_0^x S_j^x\rangle \sim  j^{-\eta}$, which indicates the 
	  correlation functions decay according to  power laws with the critical exponents (a) $\eta_1 \simeq 0.995 $ and (c) $\eta_2 \simeq 0.937 $.   All results are from the case where $g=0.5$ and $L=180$.}
   \end{figure*}

  \begin{center}
	{\centering \bf Supplementary Note 2: Spin dynamics without the magnetic field}
	\end{center}
	\begin{figure*}[t]
	  \includegraphics[width=14cm]{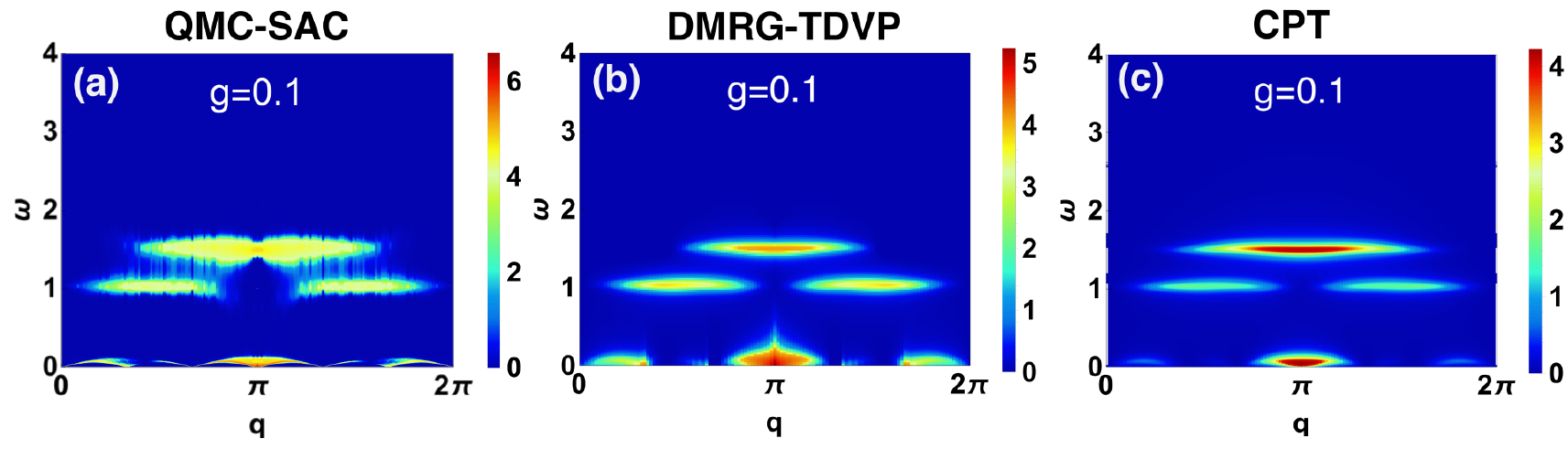}
	   \caption{\label{Sup-fig2} \textbf{Dynamic spin structure factor $\mathcal{S} (q,\omega) $  obtained from (a) QMC-SAC, (b) DMRG-TDVP and (c) CPT calculations for the trimer chain without magnetic field.}   The QMC-SAC data is sourced from our previous study \cite{cheng2022}. The  color coding of $\mathcal{S} (q,\omega)$ uses a piecewise function with the boundary value $U_0=4$. 
	   Below the boundary, the low-intensity portion is characterized by a linear mapping of the spectral
	   function to the color bar, while above the boundary a logarithmic scale is used, $U=U_0+\log_{10}[\mathcal{S}(q,\omega)]- \log_{10}(U_0)$.}
	\end{figure*}
  
	\begin{figure*}[t]
	  \includegraphics[width=14cm]{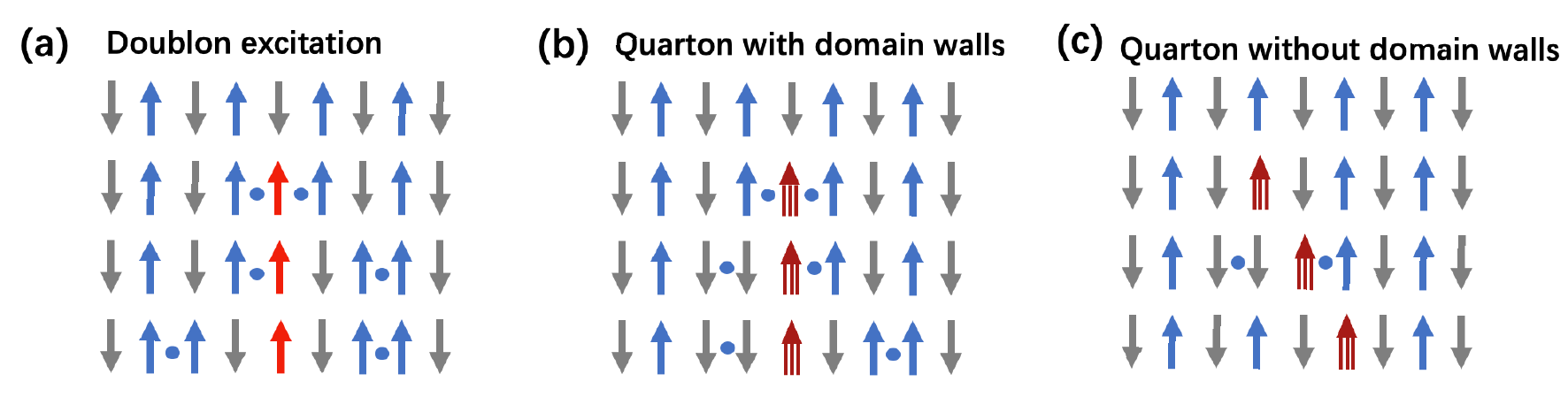}
	   \caption{\label{Sup-fig3} \textbf{Schematic representation of propagating  doublon and quarton.}  The excitation mechanism and propagation of a quasiparticle are illustrated from top to bottom. Each arrow represents  an effective spin of one trimer. (a)  A doublon excitation, in which the spinons (domain walls indicated by dots) are bound to  one excited trimer  indicated by the red color. (b)(c) Quartons with  and without domain walls for $\Delta M =1$. The dark red arrow represents the second excited  state of single trimer with $S^z =3/2$. }
	\end{figure*}

	\begin{figure*}[t]
	  \includegraphics[width=14cm]{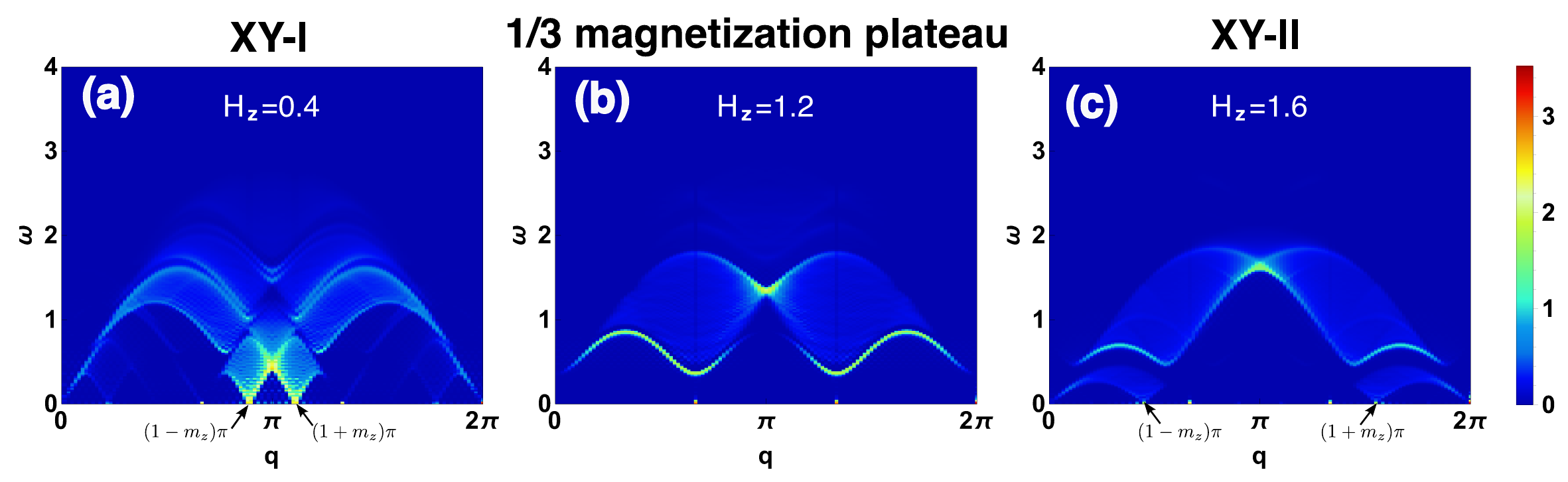}
	   \caption{\label{Sup-fig4} \textbf{Dynamic spin structure factor $\mathcal{S}^{zz}  (q,\omega) $  obtained from DMRG-TDVP calculations for different phases.}  $\mathcal{S}^{zz}  (q,\omega) $ in  (a) XY-I phase, (b) $1/3 $ magnetization  plateau phase, and (c) XY-II phase. All results are from the case where $g=0.8$ and $L=120$. The  color coding of $\mathcal{S}^{zz} (q,\omega)$ uses a piecewise function with
	   the boundary value $U_0=2$. 
	   Below the boundary, the low-intensity portion is characterized by a linear mapping of the spectral
	   function to the color bar, while above the boundary a logarithmic scale is used, $U=U_0+\log_{10}[\mathcal{S}^{zz}(q,\omega)]- \log_{10}(U_0)$.}
	\end{figure*}

  In our previous investigation \cite{cheng2022}, we explored the  spin dynamics of a trimer chain in the absence of a magnetic field using quantum Monte Carlo with subsequent numerical analytic continuation (QMC-SAC). We demonstrated that small intertrimer interactions give rise to distinct types of collective excitations associated with the internal trimer excitations. Particularly in the intermediate-energy and high-energy regimes,  the presence of  doublons and quartons was revealed through the QMC-SAC calculations and theoretical analysis. These findings were corroborated by the inelastic neutron scattering  measurements on \ce{Na_2Cu_3Ge_4O_12} \cite{bera2022emergent}, where the trimer chain is subject to the next-nearest neighbor intratrimer exchange couplings $J_3$, as depicted in  Fig.6(a) of main text. The $J_3$ interaction induces  frustration within each trimer, which may lead  to a negative sign problem  in the QMC-SAC calculations. Therefore, we mainly employ the  density matrix renormalization group and time-dependent variational principle (DMRG-TDVP)  to investigate the spin dynamics of trimer systems under a magnetic field. Supplementary Fig.~\ref{Sup-fig2} presents the results of spin dynamics $\mathcal{S} (q,\omega) = 3 \mathcal{S}^{xx} (q,\omega)  $ for $g=0.1$  obtained through the QMC-SAC, DMRG-TDVP and CPT calculations, considering   the trimer chain without  a magnetic field, which preservs the SU(2) symmetry. All three methods successfully reveal
  the low-energy continuum, as well as intermediate-energy and high-energy excitations corresponding to the spinons, doublons, and quartons, respectively. Only minor differences are present in certain spectral details. For example,  the QMC-SAC calculation effectively characterized the spinon continuum with high resolution in the low-energy regime, while the DMRG-TDVP and CPT calculations distinctly separated the doublons and quartons in the intermediate-energy and high-energy regimes.   These methods offer reliable results for exploring the spin dynamics of the trimer chain.
  
  In order to enhance the understanding of the propagation of doublons and quartons, we provide a physical  elucidation in Supplementary Fig.~\ref{Sup-fig3}. We first consider the  excitation of a doublon, as depicted in Supplementary Fig.~\ref{Sup-fig3}(a), one trimer is excited to its second-excited doublet,  resulting in   a flipped effective spin.  This excitation is characterized by $|\Delta M| = 1$  and  the formation of two domain walls.  These mobile domain walls, referred to as spinons, while not entirely free, remain tethered to the persistent central doublet.
   Consequently,  the central doublet propagates through the system, accompanied  by spinons, which gives rise to numerous internal modes within these composite excitations, thereby generating a  band of finite width. Next, we examine to the quarton, which presents several possibilities for creating an excitation with $|\Delta M| = 1$ based on the trimer excitations. For simplicity, we consider two cases involving  the $S^z = 3/2$ states to illustrate the propagation of quartons with and without domain walls.  In  Supplementary Fig.~\ref{Sup-fig3}(b),  an effective spin $S^z=-1/2$ is replaced by the $S^z = 3/2$ state, then an excitation $|\Delta M| = 2$  is generated. To achieve the excitation $|\Delta M| = 1$,  it is necessary to flip one of the neighboring $S^z=1/2$ spins downward, which creates a domain wall. Additionally, on the opposite  side of the excited trimer, a domain wall may also propagate outward. Thus,  the central  excited state  propagates throughout the system, enveloped  by two domain walls.  In  Supplementary Fig.~\ref{Sup-fig3}(c),   the replacement of an effective spin  $S^z=1/2$ with the $S^z = 3/2$ state leads to  the presence of  two domain walls, which subsequently disappear during the propagation of the central  excited state. Then a quarton propagates without domain walls.  If we restore spin-rotation symmetry and explore alternative scenarios,   quartons can emerge through analogous mechanisms. For further insights into the dynamics of doublons and quartons, we refer the interested reader to the Ref.~\cite{cheng2022}.

	\begin{center}
	  {\centering \bf Supplementary Note 3: Longitudinal excitation spectrum}
	  \end{center}

	  Let us examine the longitudinal excitation spectrum $\mathcal{S}^{zz}  (q,\omega) $ at  $g=0.8$ for the XY-I phase, the $1/3$ magnetic plateau phase, and the XY-II phase. As shown in Supplementary  Fig.~\ref{Sup-fig4},  the excitations  in  two XY phases are gapless, whereas  they are gapped in the $1/3$ magnetic plateau phase. In the ferromagnetic phase, where all spins are polarized in the $z$ direction, the longitudinal excitation spectrum is nonexistent. In both the XY-I and XY-II phases, the intraband zero-energy excitations correspond to the longitudinal  fluctuations in the framework of spinless fermion~\cite{takayoshi2023phase}. 
	  The incommensurability observed in the spin dynamics  is evidenced by  the splitting of   the excitation bands. In the XY-I phase, see Supplementary  Fig.~\ref{Sup-fig4}(a), the 
	  longitudinal excitations preserve  the total particle number without altering  the magnetization of the ground state, which indicates that the 
	  incommensurate fluctuations approaching  zero energy at $q=(1\pm m_z)\pi$ with $m_z=0.1111$ being the  magnetization normalized by its saturation value. In the XY-II phase, see Supplementary  Fig.~\ref{Sup-fig4}(c), the 
	  incommensurate fluctuations attain a zero energy state at $q=(1\pm m_z)\pi$ where $m_z = 0.5697$.

	  \begin{center}
		{\centering \bf Supplementary Note 4: Effects of intertrimer interaction on the spin dynamics }
		\end{center}
  
\begin{figure*}[t]
	  \includegraphics[width=18cm]{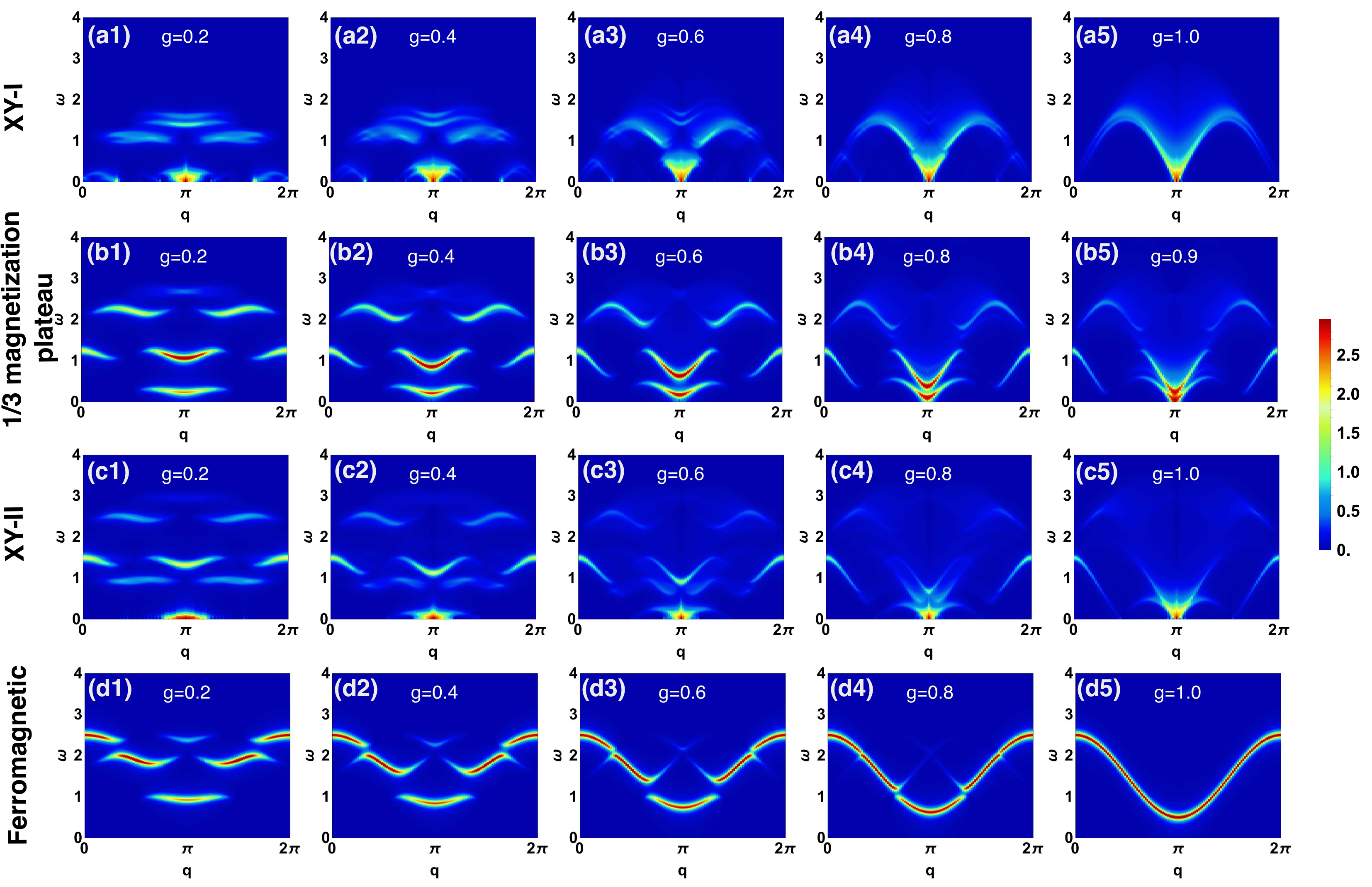}
	   \caption{\label{Sup-fig5} \textbf{ $\mathcal{S}^{xx}  (q,\omega) $  obtained from DMRG-TDVP calculations for different phases.}   $\mathcal{S}^{xx}  (q,\omega) $ in (a1)-(a5) XY-I phase ($H_z =0.1$), (b1)-(b5) $1/3 $ magnetization  plateau phase ($H_z =1. 25$), (c1)-(c5) XY-II phase, and  (d1)-(d5) ferromagnetic phase ($H_z =2.0$) for different intertrimer interaction $g$.  All results are from the case where $L=120$. Particularly,  we choose $g=0.9$ in (b5) for the reason that $H_z =1.25, g=1.0$ is a critical point. The  color coding of $\mathcal{S}^{xx} (q,\omega)$ uses a piecewise function with  the boundary value $U_0=2$.}
	\end{figure*}

  In the main text, we have discussed the effects of a magnetic field on the spin dynamics by considering  a fixed value of $g$, how the spectra evolve with varying $g$ will provide  further insights into the spin dynamics of the trimer chain under the influence of a magnetic field.  Supplementary Fig.~\ref{Sup-fig5} presents the $\mathcal{S}^{xx}  (q,\omega) $ for four distinct phases as  $g$ is varied. Within the XY-I phase, as illustrated in Supplementary Fig.~\ref{Sup-fig5} (a1)-(a5), the spectra remain gapless. A low value of  $g$ induces the clear separation of spectra corresponding to different energy levels,   which includes the presence of doublons, upper quartons, and lower quartons, as  discussed in the main text.   As $g$ increases, the  intermediate-energy and high-energy spectra gradually merge, and eventually forming a continuum with the low-energy spectrum. When $g=1$, the trimer chain becomes the  Heisenberg XXX model, a little magnetic field ($H_z=0.1$) induces an incommensurate order leading to a slight  shift at the lower boundary of the two-spinon continuum. In the $1/3 $ magnetization  plateau phase, the gap at $q=\pi$ diminishes with as $g$ increases.  The  low-energy   reduced spin wave,  intermediate-energy doublons, and high-energy quartons gradually lose their  identity and merge as $g$ approaches $1$.
  Gapless excitations and the merging of spectra are also observed in the XY-II phase, but there are some distinctions compared to the  XY-I phase. For example, when $g=0.2$, a new energy band emerges near $\omega =1$ that has no trail in other three phases.  In the ferromagnetic phase, the gap at $q=\pi$ 
  decreases with increasing $g$. The excitation spectra continue to be characterized  by the spin waves, but exhibit two energy levels for the same wave vector at the  edges of the Brillouin
  zone  due to the periodic potential  arising from the trimerized interaction. When $g=1$, the  system restores the translational  invariance, and its excitation spectrum becomes a single one  described by a cosine function of momentum.
  
  \begin{center}
	{\centering \bf Supplementary Note 5: Exact diagonalization  results}
	\end{center}

	\begin{figure*}[t]
	  \includegraphics[width=12cm]{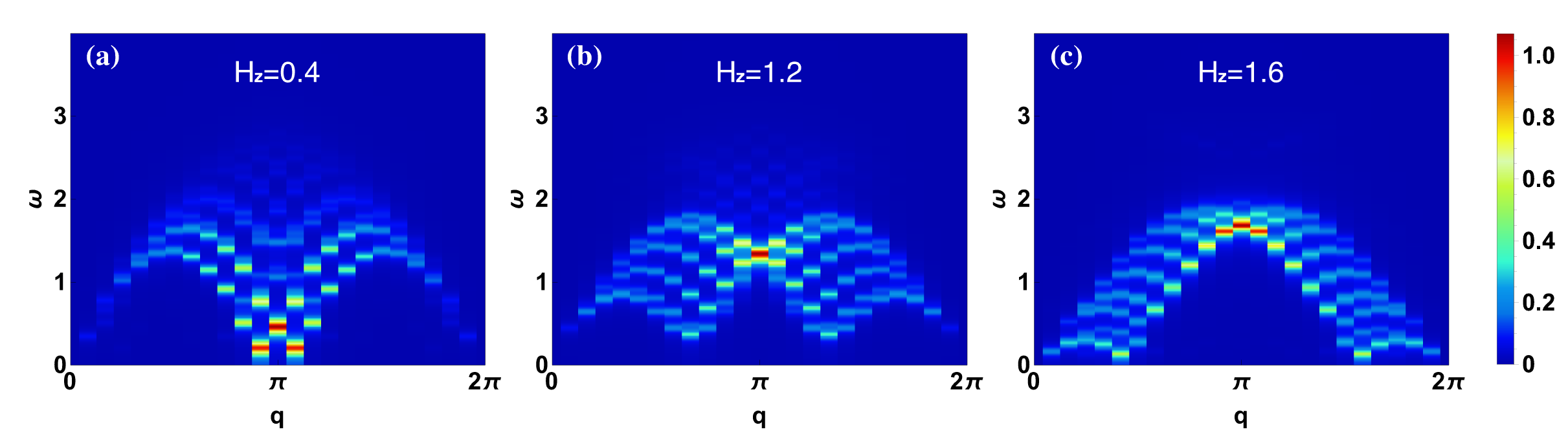}
	 \caption{\label{szz-ed} \textbf{ $\mathcal{S}^{zz}  (q,\omega) $  obtained from ED calculations for different phases.}  $\mathcal{S}^{zz}  (q,\omega) $ in (a) XY-I phase, (b) $1/3 $ magnetization  plateau phase, and (c) XY-II phase. All results are from the case where $g=0.8$ and $L=24$. The  color coding of $\mathcal{S}^{zz} (q,\omega)$ uses a piecewise function with  the boundary value $U_0=0.2$.}
  \end{figure*}

  \begin{figure*}[t]
	 \includegraphics[width=16cm]{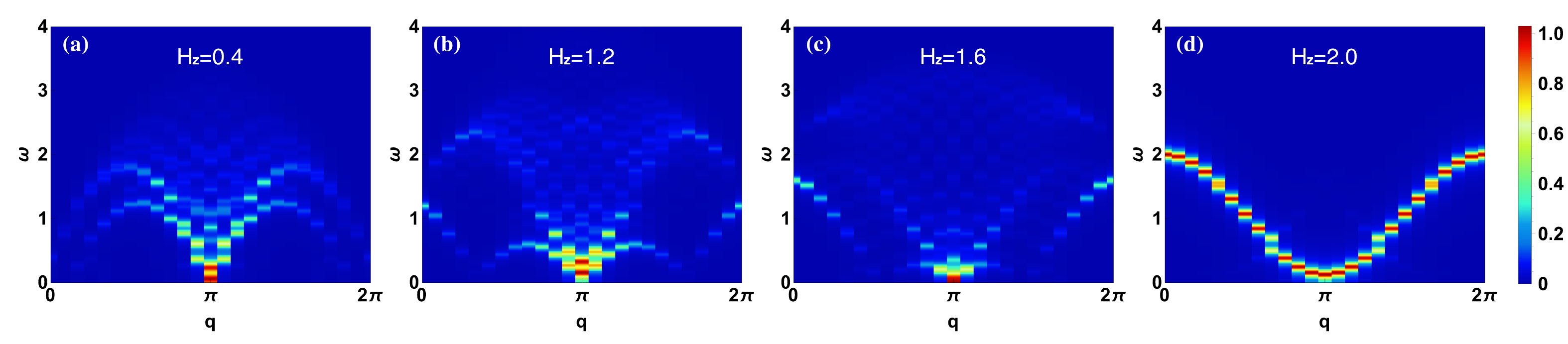}
	 \caption{\label{sxx-g0.8-ED} \textbf{$\mathcal{S}^{xx}  (q,\omega) $  obtained from ED calculations for different phases.}  $\mathcal{S}^{xx}  (q,\omega) $ in  (a) XY-I phase, (b) $1/3 $ magnetization  plateau phase, (c) XY-II phase, and (d) Ferromagnetic phase. All results are from the case where $g=0.8$ and $L=24$. The  color coding of $\mathcal{S}^{xx} (q,\omega)$ uses a piecewise function with the boundary value $U_0=0.2$.}
  \end{figure*}
  \begin{figure*}[t]
	 \includegraphics[width=16cm]{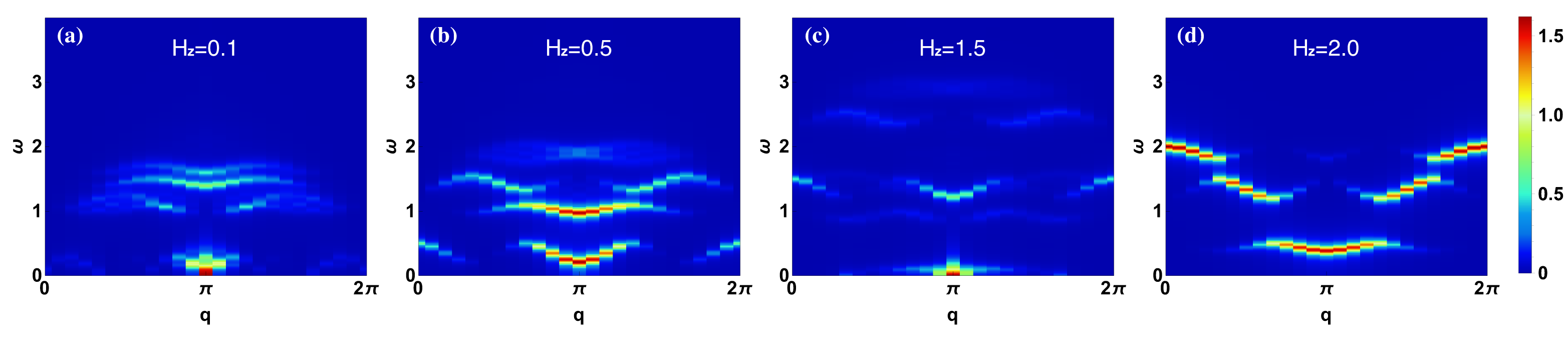}
	 \caption{\label{sxx-g0.3-ED}  \textbf{ $\mathcal{S}^{xx}  (q,\omega) $  obtained from CPT and DMRG-TDVP calculations for different phases with weak intertrimer interaction.}  $\mathcal{S}^{xx}  (q,\omega) $ in  (a) XY-I phase, (b) $1/3 $ magnetization  plateau phase, (c) XY-II phase, and (d)Ferromagnetic phase. All results are from the case where $g=0.3$ and  $L=24$. The  color coding of $\mathcal{S}^{xx} (q,\omega)$ uses a piecewise function with the boundary value $U_0=0.2$.}
  \end{figure*}

	In addition to the DMRG-TDVP and CPT calculations, we have also applied the exact diagonalization (ED) method to study the spin dynamics. The ED method serves as a fundamental and straightforward approach for calculating the eigenenergies and eigenstates of a spin models with small sizes, which is essential for  analyzing  quantum phase transitions and magnetic excitations in spin systems. In the main text,    the quantum critical points of Fig.2(b) and  all results presented in Fig.(4)  are derived from the ED calculations. Initially,  we perform ED calculations to obtain reliable results for further investigation, as these calculations are computationally efficient and require minimal resources. When addressing larger system sizes, we employ symmetries to block diagonalize the Hamiltonian, thereby reducing computational time and memory usage. Nevertheless,   the  finite-size effects remain significant. To gain more reliable insights into the thermodynamic limit, we employ advanced numerical methods, such as quantum Monte Carlo, DMRG and CPT. By comparing the results obtained from these various methods,  we can draw more credible conclusions.  
	
	In  Supplementary Fig.~\ref{szz-ed}, the longitudinal spin excitations $\mathcal{S}^{zz}  (q,\omega) $  of the XY-I, $1/3 $ magnetization  plateau, and XY-II phases are present. These findings are in excellent agreement with those presented  in Supplementary Fig.\ref{Sup-fig4}, particularly with respect to the incommensurate wave numbers at zero energy and the   characteristics of excitation spectra for the three phases. Furthermore, Supplementary Fig.~\ref{sxx-g0.8-ED} and Fig.~\ref{sxx-g0.3-ED} display the results of $\mathcal{S}^{xx}  (q,\omega) $ for  different phases,  which are consistent with the results presented  in Figs.(2) and (3) of the main text, respectively.
	

	\begin{center}
	  {\centering \bf Supplementary Note 6: Dispersion relations}
	  \end{center}
  
	In the main text, the dispersion relations provide valuable insights into the excitation mechanisms underlying various spin dynamics. When $g$ is small, the intermediate-energy and high-energy excitations are primarily localized within the trimers. To confirm the nature of these quasiparticles across different phases, we propose a methodology for deriving  their dispersion relations through the imitation of complex ground states.  These dispersion relations are consistent  with the DMRG-TDVP and CPT results on the location and band widths of excitations spectra (see Fig.3(e) and Fig.5 of main text), indicating that our understanding of the excitation remains accurate, despite the utilization of a highly simplified approximation for the ground state in our calculations. 
	
	Here, we delineate the primary  procedures for deriving the dispersion relations.  In the absence of a magnetic field, the trimer chain can be effectively described by an antiferromagnetic Heisenberg model, where the effective interaction is contingent upon the intertrimer  interaction $J_{\rm{eff}}=4J_2/9$ \cite{cheng2022}.  Given  the weak intertrimer interaction and the doubly degenerate ground state, each trimer can be represented as  an effective spin $S=1/2$. Therefore, the assumption that  the ground-state wave function of spin chain is a product state of ground states of each trimer provides a viable  approach  to simulate the internal trimer excitations. Upon the introduction of a magnetic field,  as discussed in the main text, a quantum phase transition presents, resulting in gapless or gapped ground states in different phases. Our findings indicate that this assumption remains applicable for   analyzing excitations in the XY-I and the $1/3 $ magnetization  plateau phases. 
	In the XY-I phase, we assume that the ground-state wave function of the trimer chain is a product state of  $\left|0\right\rangle$ and $\left|1\right\rangle$, 
	\begin{equation}
		\label{ground state}
		\left| {\psi}\right\rangle_{\mathrm{g}}=\left| {0}\right\rangle_1 \left| {1}\right\rangle_2 \cdots \left| {0}\right\rangle_{N-1}  \left| {1}\right\rangle_{N},
	\end{equation}
	where $\left|0\right\rangle$ and $\left|1\right\rangle$ is the ground state and first excited state of a single trimer, respectively, as shown in Fig.(4) of main text. We  disregard interactions  and instead focus on $2^{N-1}$ degenerate ground states, while imposing the constraint on  the magnetic quantum numbers $\sum M_i = 0 $. Even if the  magnetic field induces an incommensurate ground state with minimal  magnetization, the aforementioned assumption continues to hold.  It is important to emphasize that our objective is to investigate excitations above the $2^N$-fold degenerate ground-state manifold, rather than employing degenerate  perturbation theory. In this study,  we do not engage in a formal perturbation expansion; rather, we opt to construct intuitive variational states that encompass the internal excitations of a single trimer. 
	
\begin{figure*}[t]
		\includegraphics[width=16cm]{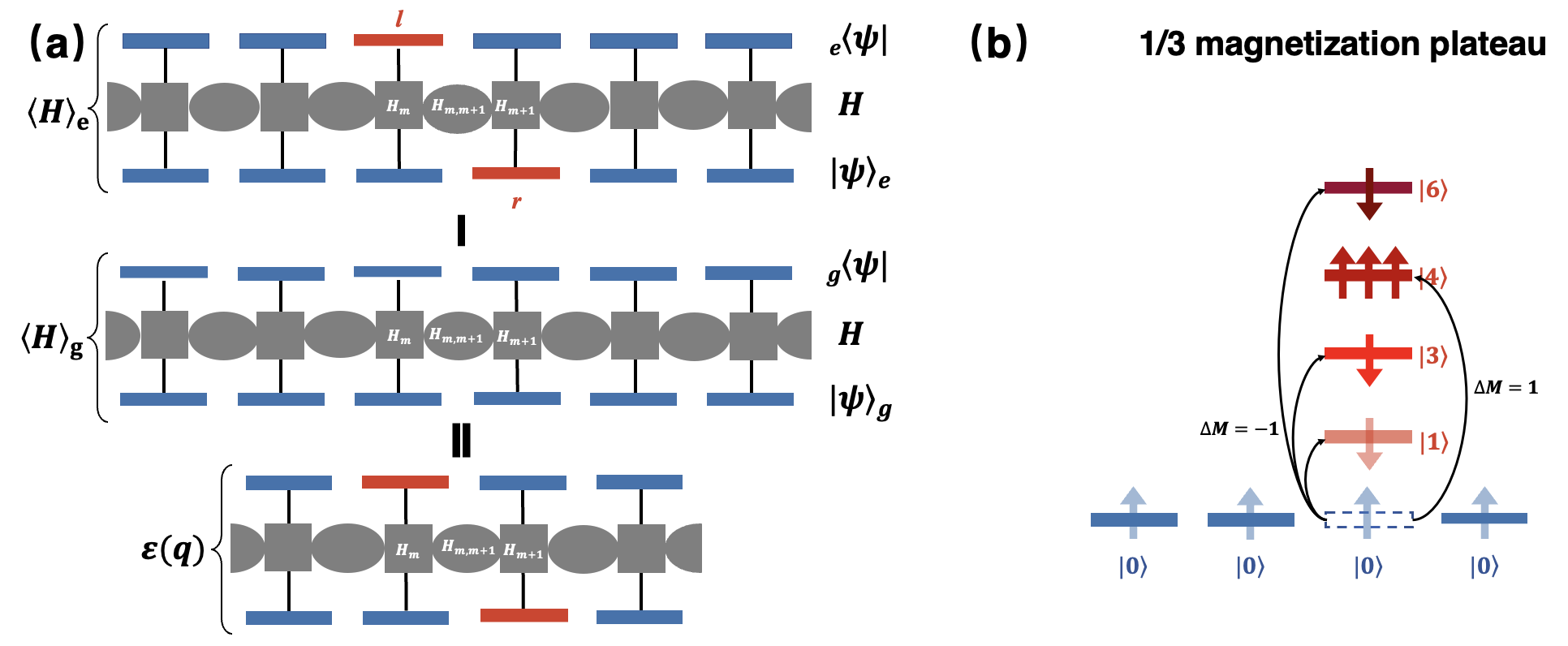}
		\caption{\label{dispersion-cal} \textbf{Graphical representation of the calculation of the dispersion relation
		$\epsilon(q)=\left\langle H \right\rangle_{\rm{e}} - \left\langle H \right\rangle_{\rm{g}}$.} (a) The  trimer eigenstates  are shown as darker blue (ground states)
		and  red (excited states). With these states, the  excitations with $|\Delta M|=1$ originate from a trimer ground state
		$\left|0\right\rangle^{1}_r$ on the trimer located at $r$ when excited to one of  $\left|1\right\rangle_r$, $\left|3\right\rangle_r$, $\left|4\right\rangle_r$, and $\left|6\right\rangle_r$ (and in the corresponding bra states we use the site index $l$ instead of $r$). The excitations are given momentum $q$, and the matrix elements contributing
		to the dispersion relation are indicated. (b) The construction of ground state and excitation mechanism
	in the  $1/3 $ magnetization  plateau phase.}
	\end{figure*}

	For the intermediate-energy excitations, we choose the $r$th trimer to be excited from $\left|0\right\rangle$ to $\left|3\right\rangle$ with $\Delta M =-1$ or  from $\left|1\right\rangle$ to $\left|2\right\rangle$ with $\Delta M =1$. The corresponding  excited wave function is then given by 
	
	\begin{equation}
		\left| {\psi}\right\rangle_{\mathrm{e}}^r=\left| {0}\right\rangle_1 \left| {1}\right\rangle_2 \cdots  \left| {3}\right\rangle_r\cdots \left| {0}\right\rangle_{N-1} \left| {1}\right\rangle_{N}.
	\end{equation}
	To impart momentum to this excitation, we perform a Fourier transformation on the excited state, resulting in 
	\begin{equation}
		\left| {\psi }\right\rangle_{\mathrm{e}}^q=\frac{1}{\sqrt{N}} \sum_{r=1}^{N} \mathrm{e}^{-\mathrm{i} \emph{q} \emph{r}} \left| {\psi}\right\rangle_{\mathrm{e}}^r.
	\end{equation}
	Next, we proceed to calculate the expectation values of the Hamiltonian in both the ground state and the first excited trimer momentum state to  derive the dispersion relations associated with  the intermediate-energy excitations within the reduced Hilbert space,
	\begin{eqnarray}
		\epsilon(q)&=&\left\langle H \right\rangle_{\mathrm{e}} - \left\langle H \right\rangle_{\mathrm{g}} \nonumber \\
		&=& ^{q}_{\mathrm{e}}\left\langle \psi\right|H \left| \psi\right\rangle^{q}_{\mathrm{e}} -^{q}_{\mathrm{g}}\left\langle \psi\right|H \left| \psi\right\rangle^{q}_{\mathrm{g}}.
	\end{eqnarray}
	Thus, the dispersion relations corresponding to the intermediate-energy doublon are given by,
	\begin{equation}
		\epsilon_{\mathrm{D}}^{\mathrm{red}} (q)=\left\{
		\begin{split}
	& -  \frac{1}{3} g \cos{q}+\mathbb{E}_1 -\mathbb{E}_0 + H_z - \frac{1}{9}g,\\
	& - \frac{2}{9} g \cos{q}+\mathbb{E}_1 -\mathbb{E}_0 + H_z,\\
	& - \frac{2}{9} g \cos{q}+\mathbb{E}_1 -\mathbb{E}_0 + H_z + \frac{2}{9}g,\\
	& - \frac{2}{9} g \cos{q}+\mathbb{E}_1-\mathbb{E}_0  + H_z + \frac{2}{9}g, 
		\end{split}
	 \right.
	\end{equation}
	which are independent of the length of the spin chain. $\mathbb{E}_0$ and $\mathbb{E}_1$ are the ground-state and first excited-state energies of one single trimer in the absence of magnetic field, respectively. 
	By substituting $q$  with $3q$, we can derive the unfolded dispersion relations  across the entire Brillouin zone, as shown in Eq.(11) of main text.
	To illustrate the computational  process, we present Supplementary Fig.~\ref{dispersion-cal}, which demonstrates that only $4$ trimers are necessary to obtain all the dispersion relations. 
	For the high-energy excitations, the $r$-th trimer is excited 
	from $\left|0\right\rangle$ to $\left|4\right\rangle$ with $\Delta M =1$ or  from $\left|0\right\rangle$ to $\left|6\right\rangle$ with $\Delta M =-1$. Similar calculations can be performed to obtain the dispersion relations of the high-energy modes. However, the magnetic field results in the splitting of the  high-energy spectra into two branches, attributable to the varying  spin quantum numbers $\Delta M = \pm 1$.  These branches are  referred to as  the upper quarton and lower quarton (see Eq.(12) and Eq.(13) of main text), respectively. 
	
	In the   $1/3 $ magnetization  plateau phase, the gapped ground state facilitates  the calculations. As depicted in  Supplementary Fig.~\ref{dispersion-cal}(b),  the  ground state is constructed from a product of polarized trimers (as the effective spins $S_{\rm{eff}}=1/2$).  The low-energy excitation arise from the flipping of an effective spin, akin to the formation of spin wave,  but within a reduced Hilbert space. Therefore, the  excitation from $\left|0\right\rangle$ to $\left|1\right\rangle$ with $\Delta M =-1$ is characterized by the reduced spin wave.

	Additional excitations also emerge from the internal trimer excitations, such as  $\left|0\right\rangle \rightarrow \left|3\right\rangle$ with $\Delta M =-1$, $\left|0\right\rangle \rightarrow \left|4\right\rangle$ with $\Delta M =1$, and $\left|0\right\rangle \rightarrow \left|6\right\rangle$ with $\Delta M =-1$. Each excitation  possesses a singular  dispersion relation that aligns well with the excitation spectrum. In the high-energy regime, the presence of a continuum with  weak intensity may  originate from the fractional spinons, as discussed in our previous study \cite{cheng2022}. Furthermore, in Supplementary Fig.~\ref{large-g-dispersion},  the gap between the ground state and the first excited state provides effective protection for the aforementioned excitations. Even when the intertrimer interaction increases to $g=0.8$, the dispersion relations continue to effectively capture the internal trimer excitations.  An increase in the intertrimer interaction $g$ leads to the merging of these excitation spectra, leading to  the formation of a continuum.

	\begin{figure*}[t]
		\includegraphics[width=12cm]{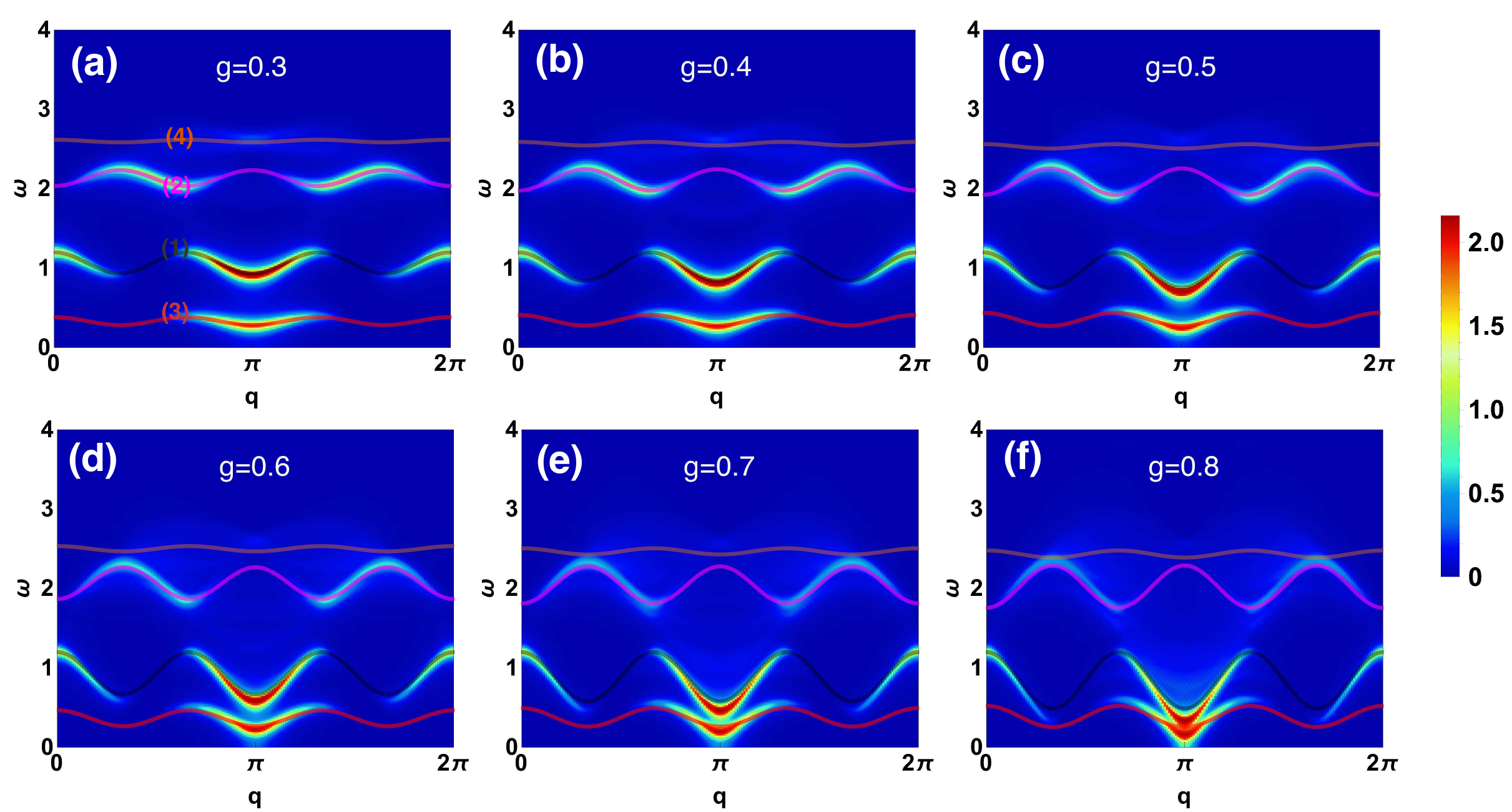}
		\caption{ \label{large-g-dispersion}  \textbf{ $\mathcal{S}^{xx}  (q,\omega) $ in  $1/3 $ magnetization  plateau phase for different $g$.} All results are obtained by  DMRG-TDVP calculations  for $L=120$, $H_z=1.2$, and the  color coding of $\mathcal{S}^{xx} (q,\omega)$ uses a piecewise function with the boundary value $U_0=2$. The dispersion lines with colors and numbers are corresponding to the different localized excitations in a single trimer. (1)(2)(4) are the excitations from $\left| 0 \right\rangle \rightarrow \left| 1 \right\rangle$, $\left| 0 \right\rangle \rightarrow \left| 3 \right\rangle$ and $\left| 0 \right\rangle \rightarrow \left| 6 \right\rangle$ with $\Delta M =-1$, respectively. (3) is the  excitations from $\left| 0 \right\rangle \rightarrow \left| 4 \right\rangle$ with $\Delta M =1$.}
	\end{figure*}

\end{document}